\documentclass[preprint]{aastex}

\usepackage{amsmath}
\usepackage{graphicx}
\usepackage{wasysym}

\begin{document}

\title{On the stability of dust orbits in mean motion resonances
with considered perturbation from an interstellar wind}

\author{Pavol P\'{a}stor}
\affil{Tekov Observatory, Sokolovsk\'{a} 21, 934~01 Levice, Slovak Republic}
\email{pavol.pastor@hvezdarenlevice.sk,~pastor.pavol@gmail.com}

\begin{abstract}
Circumstellar dust particles can be captured in a mean motion
resonance with a planet and simultaneously be affected by non-gravitational
effects. It is possible to describe the secular variations of a particle
orbit in the mean motion resonance analytically using averaged
resonant equations. We derive the averaged resonant equations from
the equations of motion in near-canonical form. The secular variations
of the particle orbit depending on the orientation of the orbit in space
are taken into account. The averaged resonant equations can be
derived/confirmed also from Lagrange's planetary equations.
We apply the derived theory to the case when the non-gravitational effects
are the Poynting--Robertson effect, the radial stellar wind, and an
interstellar wind. The analytical and numerical results obtained are
in excellent agreement. We found that the types of orbits correspond to
libration centers of the conservative problem. The averaged resonant
equations can lead to a system of equations which hold for
stationary points in a subset of resonant variables.
Using this system we show analytically that for the considered
non-gravitational effects, all stationary points should correspond
to orbits which are stationary in interplanetary space after
an averaging over a synodic period. In an exact resonance,
the stationary orbits are stable. The stability is achieved
by a periodic repetition of the evolution during the synodic period.
Numerical solutions of this system show that there are no stationary
orbits for either the exact or non-exact resonances.
\end{abstract}


\section{Introduction}
\label{sec:intro}

A small body orbiting a star with a planet is affected by the planet's
gravitational field. When the orbital periods of the body and the planet have
a ratio of small natural numbers, mean motion resonances (MMRs) can occur.
If the body is captured in an MMR, then variations of the semimajor axis
are balanced by the resonant interaction with the planet's field of gravity.
Examples of MMRs in the Solar system are numerous. We can mention the MMRs
of asteroids with Jupiter, which have resulted in the formation of the Kirkwood
gaps in the main asteroid belt \citep{kirkwood}, MMRs
of particles in Saturn's rings with its moons, MMRs
of the Edgeworth--Kuiper belt objects with Neptune, MMRs
of dust particles with the Earth \citep{ring,IRAS,COBE}, and so on.
Following the observational confirmation of dust particles being captured
in MMRs with the Earth by satellites {\it IRAS} \citep{IRAS} and
{\it COBE} \citep{COBE}, unknown properties of orbital evolution for
dust particles captured in MMRs have become an interesting topic of research.

If the planet moves in a circular orbit around the star and the mass
of the particle is negligible in comparison with the mass of the star
and also in comparison with the mass of the planet, we have a special
gravitational problem of three bodies. This gravitational problem
is called the circular restricted three-body problem (CR3BP) in celestial
mechanics.

Unlike larger bodies such as the planets or moons, dust particles can
be significantly affected by weak non-gravitational forces. The acceleration
of the particle caused by electromagnetic radiation described by
the Poynting--Robertson (PR) effect \citep{poynting,robertson,icarus}
was used within the framework of the planar CR3BP by \citet{FM2} in order
to obtain stationary solutions for dust particles captured in MMRs.
They found that for all dust particles captured in a given exterior
resonance (the orbital period of the particle in the exterior resonance is
larger than the planet's orbital period), the eccentricities of the orbits
approach a constant value, which they called the ``universal
eccentricity''. Their derivation starts from a near-canonical
form of the equations of motion in Cartesian coordinates with
the PR effect used as a dissipative external acceleration
(for dissipative accelerations in the CR3BP
and periodic orbits see also \citealt{margheri}).
Another near-canonical approach was used in \citet{gomes95}
in order to determine which non-gravitational forces (with assumed
behavior leading to the universal eccentricity) give a stable,
and which an unstable, capture in an exterior resonance within
the framework of the planar CR3BP. The author began from
a near-canonical system of equations written in Delaunay variables.
The secular time derivative of the eccentricity for a dust particle
captured in an MMR in the planar CR3BP under the action
of arbitrary non-gravitational effects can be obtained using
the method presented in \citet{gomes95}. Within the framework
of the CR3BP, there exists a constant, referred to as
the Jacobi constant, which depends on the position and velocity
of the particle and the position of the planet. Under the assumptions
that the particle is far enough from the planet and that the mass
of the planet is negligible compared to the mass of the star,
the Jacobi constant reduces to the so called Tisserand parameter.
The Tisserand parameter depends only on the Keplerian orbital elements
of the particle and the planet. For the PR effect, the universal
eccentricity was confirmed using a different derivation starting from
the secular time derivative of the Tisserand parameter by \citet{LZ1997}.
In \citet{LZ1997}, also, a general relation between the secular time
derivatives of the eccentricity and inclination during the orbital evolution
of a dust particle captured in an MMR in the CR3BP with the PR
effect and a radial stellar wind (for the wind, see \citealt{gustafson,covsw})
has been found. In \citet{gomes97} are derived the secular time derivatives
of the eccentricity and inclination from both a Jacobi constant
approach and another approach explicitly using the disturbing function.
The equations are deduced depending on which resonance captured
the dust particle. The influence of the PR effect and a non-radial
stellar wind \citep{covsw} on the secular evolution of the eccentricity
for a dust particle captured in an MMR within the framework of the planar
CR3BP was analytically investigated in \citet{AA2} and \cite{AA3}.
The secular evolutions of the eccentricity and the argument of pericenter
of a dust particle captured in an MMR in the planar circular and elliptical
restricted three-body problem with the PR effect were numerically
investigated in \citet{CMDA}.

The influence of an interstellar wind on the motion
of dust particles was already mentioned in \citet{comet}.
Observations of debris disks around stars show that the interaction
of the interstellar wind with the dust particles of the disk can change
the shape of the debris disks \citep{hines,maness,debes,buenzli,rodigas}.
From among only 16 debris disks resolved in scattered light \citep{golimowski},
as many as three disks show a changed morphology caused by the interaction
with the surrounding interstellar matter. The acceleration acting on
a spherical body moving through an interstellar gas was published in
\citet{baines}. This acceleration was used in the calculation of the secular
time derivatives of the Keplerian orbital elements caused by an interstellar
gas flow (IGF) for an arbitrary orientation of the orbit in \citet{flow} and
\citet{dyncd}. The secular decrease of the semimajor axis for all initial
conditions was found. This result was confirmed analytically by \citet{bera}
and numerically by \citet{marthe} and \citet{marzari}. \citet{bera} used
an orbit-averaged Hamiltonian approach to determine the orbital evolution
of a dust particle in a Keplerian potential subject to an additional
constant force. If the speed of the IGF is much greater than the speed
of the dust grain and also much greater than the mean thermal speed
of the gas in the flow, then the acceleration caused by
the IGF reduces to a constant vector. Therefore, the problem
which they solved can be applied to the orbital
motion of a dust particle under the action of an IGF.
The secular orbital motion in this case can be completely
solved analytically \citep{fgf}.

In \citet{mmrflow}, the results of \citet{dyncd} were used in order to obtain
the secular time derivative of the orbit eccentricity for a dust particle
captured in an MMR within the framework of the planar CR3BP with
the PR effect, a radial stellar wind, and an IGF. The agreement between
the analytical and numerical results in \citet{mmrflow} was excellent
for the IGF in the Solar system. The results show that two sets of orbital
evolutions divided by a strong boundary in the initial conditions exist
for the exterior 2/1 resonance. Also, some kind of stabilization was found
in one of these sets. These results motivated us to determine some
of the basic properties of the orbital evolution of a dust particle captured
in an MMR in the planar CR3BP with the PR effect, a radial stellar wind,
and an IGF, using the improved methods from \cite{FM1}.

\section{Averaged resonant equations}
\label{sec:average}

We will consider the motion of a homogeneous spherical dust particle
in the vicinity of a star with one planet moving in a circular orbit (CR3BP).
Non-gravitational effects will be added to the CR3BP framework.
The secular variations of the particle orbit caused by these non-gravitational
effects can depend on the orientation of the orbit in space.
Taking into account this dependence, we will improve the method presented in
\citet{FM1} for the study of problems which involve non-gravitational
effects without a spherical symmetry around the star.

The equation of motion can be written in the form
\begin{equation}\label{grad}
\frac{d \vec{v}}{dt} = \frac{\partial F}{\partial \vec{r}} + \vec{Y} ~,
\end{equation}
where $\vec{v}$ $=$ $d \vec{r} / dt$ is the velocity of the dust particle,
$\vec{r}$ is the position vector of the dust particle with respect
to the star, $F$ is the Hamiltonian of the conservative system, and
$\vec{Y}$ is the non-conservative part of the acceleration.
If we add to Eq. (\ref{grad}) equations for the velocity in Cartesian
coordinates, we obtain a system of equations in the near-canonical form:
\begin{equation}\label{canonical}
\frac{d \vec{r}}{dt} = - \frac{\partial F}{\partial \vec{v}} ~, ~~
\frac{d \vec{v}}{dt} = \frac{\partial F}{\partial \vec{r}} + \vec{Y} ~.
\end{equation}

Following \citet{FM1}, we will use canonical variables derived from
the Delaunay variables in the extended phase space. The canonical
variables are $l_{j}$ $=$ ($\lambda$, $\tilde{\omega}$, $t$) and
$L_{j}$ $=$ ($L$, $G-L$, $\Lambda$), where $\lambda$ is the mean longitude,
$\tilde{\omega}$ is the longitude of pericenter,
$L$ $=$ $\sqrt{\mu ( 1 - \beta ) a}$, $G$ $=$ $L$ $\sqrt{1 - e^{2}}$,
and $\Lambda$ is the canonical momentum that is conjugate to time.
In the canonical variables $\mu$ $=$ $G_{0} M_{\star}$, $G_{0}$
is the gravitational constant, $M_{\star}$ is the mass
of the star, $a$ is the semimajor axis of the particle
orbit, and $e$ is the eccentricity of the particle orbit.
From the non-gravitational effects, the electromagnetic radiation
of the star in the form of the Poynting--Robertson (PR) effect is most
often considered. The parameter $\beta$ is defined as the ratio
between the electromagnetic radiation pressure force and the gravitational
force between the star and the particle at rest with respect to the star
\begin{equation}\label{beta}
\beta = \frac{3 L_{\star} \bar{Q}'_{\text{pr}}}{16 \pi c \mu R_{\text{d}}
\varrho} ~.
\end{equation}
Here, $L_{\star}$ is the stellar luminosity, $\bar{Q}'_{\text{pr}}$ is
the dimensionless efficiency factor for the radiation pressure averaged
over the stellar spectrum and calculated for the radial direction
($\bar{Q}'_{\text{pr}}$ $=$ 1 for a perfectly absorbing sphere),
$c$ is the speed of light in vacuum, and $R_{\text{d}}$ is
the radius of the dust particle with mass density $\varrho$.
The radial term not depending on the particle velocity in the PR effect
(Eq. \ref{PR}) can be added to the stellar gravity because its net effect
is only to reduce the mass of the star to $M_{\star} ( 1 - \beta )$.
Therefore, we use $L$ $=$ $\sqrt{\mu ( 1 - \beta ) a}$.
Using the Brouwer--Hori theorem \citep{BH}, we can transform
Eqs. (\ref{canonical}) into a new system of near-canonical equations
\begin{equation}\label{first}
\frac{d l_{j}}{dt} = - \frac{\partial F}{\partial L_{j}} + Q_{j} ~, ~~
\frac{d L_{j}}{dt} = \frac{\partial F}{\partial l_{j}} + P_{j} ~,
\end{equation}
where
\begin{equation}\label{theorem}
Q_{j} = \sum_{k = 1}^{3} Y_{k} \frac{\partial r_{k}}{\partial L_{j}} ~, ~~
P_{j} = \sum_{k = 1}^{3} Y_{k} \frac{\partial r_{k}}{\partial l_{j}}
\end{equation}
and $r_{k}$ are the Cartesian coordinates of the particle position vector
with respect to the star. A cumbersome calculation of the partial derivatives
finally yields
\begin{align}\label{PQ}
Q_{1} &= \frac{2}{L} \vec{Y} \cdot \vec{r} + ( 1 - \alpha ) Q_{2} ~,
\notag \\
Q_{2} &= \frac{a}{L e} \vec{Y}
      \left (
      \begin{array}{c}
      \alpha \cos \tilde{\omega} - e \sin E \sin \tilde{\omega}
      \\
      \alpha \sin \tilde{\omega} + e \sin E \cos \tilde{\omega}
      \end{array}
      \right ) - \frac{\alpha}{L e} P_{1} \sin E ~,
\notag \\
P_{1} &= \frac{1}{n} \vec{Y} \cdot \vec{v} ~,
\notag \\
P_{2} &= \vec{Y}
      \left (
      \begin{array}{c}
      - r_{2}
      \\
      r_{1}
      \end{array}
      \right ) - P_{1} ~,
\end{align}
where $\alpha$ $=$ $\sqrt{1 - e^{2}}$, $E$ is the eccentric anomaly,
and $n$ $=$ $\sqrt{\mu ( 1 - \beta ) / a^{3}}$ is the mean motion of
the particle. Eqs. (\ref{PQ}) differ from the equations in \citet{FM1}
due to the dependence of the partial derivatives on the orientation
of the orbit of the particle in space, which is taken into consideration.
The Hamiltonian of the conservative system in the new variables is
\begin{equation}\label{Hamiltonian}
F = \frac{\mu^{2} ( 1 - \beta )^{2}}{2 L^{2}} - \Lambda + R ~,
\end{equation}
where $R$ is the disturbing function
\begin{equation}\label{DF}
R = G_{0} M_{\text{P}} \left ( \frac{1}{\vert \vec{r} -
\vec{r}_{\text{P}} \vert} -
\frac{\vec{r} \cdot \vec{r}_{\text{P}}}{r_{\text{P}}^{3}} \right ) ~,
\end{equation}
$M_{\text{P}}$ is the mass of the planet, $\vec{r}_{\text{P}}$ is
the position vector of the planet with respect to the star, and
$r_{\text{P}}$ $=$ $\vert \vec{r}_{\text{P}} \vert$.
The subscript $\text{P}$ will be used for quantities belonging to the planet.
In order to investigate the behavior of a particle captured in an MMR
with a planet in a circular orbit, we will use the canonical
resonant variables $H$, $K$, $\sigma_{1}$, $J_{1}$, $\sigma_{2}$,
$J_{2}$
\begin{align}\label{regular}
H &= \sqrt{2 \vert L - G \vert} \sin (\psi / q - \tilde{\omega}) ~, &
K &= \sqrt{2 \vert L - G \vert} \cos (\psi / q - \tilde{\omega}) ~,
\notag \\
\sigma_{1} &= \psi / q - \tilde{\omega}_{\text{P}} ~, &
J_{1} &= G + \Lambda / n_{\text{P}} ~,
\notag \\
\sigma_{2} &= ( \lambda - \lambda_{\text{P}} ) / q ~, &
J_{2} &= ( p + q ) L + p \Lambda / n_{\text{P}} ~,
\end{align}
where $p$ and $q$ are two integers and $\psi$ $=$
$( p + q ) \lambda_{\text{P}} - p \lambda$. A second application
of the Brouwer--Hori theorem yields equations which are formally
consistent with the set of equations presented in \citet{FM1}:
\begin{align}\label{second}
\frac{dH}{dt} &= - \frac{\partial F}{\partial K} - \frac{H}{H^{2} + K^{2}}
      P_{2} + K ( s Q_{1} + Q_{2} ) ~,
\notag \\
\frac{dK}{dt} &= \frac{\partial F}{\partial H} - \frac{K}{H^{2} + K^{2}}
      P_{2} - H ( s Q_{1} + Q_{2} ) ~,
\notag \\
\frac{d \sigma_{1}}{dt} &= - \frac{\partial F}{\partial J_{1}} + s Q_{1} ~,
\notag \\
\frac{dJ_{1}}{dt} &= \frac{\partial F}{\partial \sigma_{1}} +
      P_{1} + P_{2} ~,
\notag \\
\frac{d \sigma_{2}}{dt} &= - \frac{\partial F}{\partial J_{2}} -
      \frac{1}{q} Q_{1} ~,
\notag \\
\frac{dJ_{2}}{dt} &= \frac{\partial F}{\partial \sigma_{2}} +
      ( p + q ) P_{1} ~,
\end{align}
where $s$ $=$ $p / q$. All terms in the disturbing function depending
on $\tilde{\omega}_{\text{P}}$ contain powers of $e_{1}$ as factors due
to the d'Alembert property \citep{mude}. Because we are considering only
circular planetary orbits, and $\tilde{\omega}_{\text{P}}$ is only present
in $\sigma_{1}$, we have $\partial F / \partial \sigma_{1}$ $=$ 0.
Eqs. (\ref{second}) have to be averaged over a synodic period. The change
of $\sigma_{2}$ is equal to $2 \pi$ after the synodic period.
The term $\partial F / \partial \sigma_{2}$ in
Eqs. (\ref{second}), when averaged over the synodic
period, becomes $1 / 2 \pi$ $\int_{0}^{2 \pi}$
$\partial F / \partial \sigma_{2}$ $d \sigma_{2}$ $=$
$1 / 2 \pi$ $\left [ F( 2 \pi ) - F( 0 ) \right ]$ $=$ 0
due to the periodicity of the disturbing function with this period
(during the averaging). Thus, Eqs. (\ref{second}) averaged over
the synodic period become
\begin{align}\label{averaged}
\frac{dH}{dt} &= - \frac{\partial F}{\partial K} - \frac{H}{H^{2} + K^{2}}
      \left \langle P_{2} \right \rangle +
      K ( s \left \langle Q_{1} \right \rangle +
      \left \langle Q_{2} \right \rangle ) ~,
\notag \\
\frac{dK}{dt} &= \frac{\partial F}{\partial H} - \frac{K}{H^{2} + K^{2}}
      \left \langle P_{2} \right \rangle -
      H ( s \left \langle Q_{1} \right \rangle +
      \left \langle Q_{2} \right \rangle ) ~,
\notag \\
\frac{d \sigma_{1}}{dt} &= - \frac{\partial F}{\partial J_{1}} +
      s \left \langle Q_{1} \right \rangle ~,
\notag \\
\frac{dJ_{1}}{dt} &= \left \langle P_{1} \right \rangle +
      \left \langle P_{2} \right \rangle ~,
\notag \\
\frac{d \sigma_{2}}{dt} &= - \frac{\partial F}{\partial J_{2}} -
      \frac{1}{q} \left \langle Q_{1} \right \rangle ~,
\notag \\
\frac{dJ_{2}}{dt} &= ( p + q ) \left \langle P_{1} \right \rangle ~.
\end{align}
Angle brackets for conservative quantities are omitted in Eqs. (\ref{averaged}).
In resonant problems, it is convenient to study the behavior in
the non-canonical variables ($k$, $h$) $=$ ($e \cos \sigma_{0}$,
$e \sin \sigma_{0}$), where $\sigma_{0}$ $=$ $\psi / q - \tilde{\omega}$.
$J_{1}$ can be expressed using $L$, $J_{2}$, $k$, and $h$ in the disturbing
function $R$. The number of equations can be reduced from six to five because
$dJ_{2} / dt$ can be expressed using $dL / dt$ and $dJ_{1} / dt$. The equation
for $dJ_{1} / dt$ can be replaced by an equation for $dL / dt$. A further
reduction in the number of equations is possible because $\sigma_{2}$ is
cyclic and can be ignored. Therefore, only four equations remain:
\begin{align}\label{transformed}
\frac{dk}{dt} &= \frac{\alpha}{L} \frac{\partial R}{\partial h} -
      h \frac{d \sigma_{1}}{dt} -
      \frac{k \alpha}{L ( 1 + \alpha )} \frac{dL}{dt} -
      \frac{k \alpha}{L e^{2}} \left \langle P_{2} \right \rangle -
      h \left \langle Q_{2} \right \rangle ~,
\notag \\
\frac{dh}{dt} &= - \frac{\alpha}{L} \frac{\partial R}{\partial k} +
      k \frac{d \sigma_{1}}{dt} -
      \frac{h \alpha}{L ( 1 + \alpha )} \frac{dL}{dt} -
      \frac{h \alpha}{L e^{2}} \left \langle P_{2} \right \rangle +
      k \left \langle Q_{2} \right \rangle ~,
\notag \\
\frac{dL}{dt} &= s \left ( h \frac{\partial R}{\partial k} -
      k \frac{\partial R}{\partial h} \right ) +
      \left \langle P_{1} \right \rangle ~,
\notag \\
\frac{d \sigma_{1}}{dt} &= n_{\text{P}} \frac{p + q}{q} - n s +
      \frac{2 s L}{\mu ( 1 - \beta )} \frac{\partial R}{\partial a} -
      \frac{\alpha s}{L ( 1 + \alpha )}
      \left ( k \frac{\partial R}{\partial k} +
      h \frac{\partial R}{\partial h} \right ) +
      s \left \langle Q_{1} \right \rangle ~.
\end{align}
Eqs. (\ref{transformed}) represent averaged resonant equations that
include the directional character of the non-conservative acceleration
$\vec{Y}$. It is worth mentioning that the averaged values of
the non-conservative terms $\left \langle Q_{j} \right \rangle$ and
$\left \langle P_{j} \right \rangle$ can be expressed using the averaged
values of $da / dt$, $de / dt$, $d \tilde{\omega} / dt$, and
$d \sigma_{\text{b}} / dt$ $+$ $t$ $dn / dt$ caused by
the non-gravitational effects only. The angle $\sigma_{\text{b}}$
is defined so that the mean anomaly can be computed from $M$ $=$ $n t$
$+$ $\sigma_{\text{b}}$ \citep{fund}. On the left-hand sides of
Eqs. (\ref{transformed}), we can use Lagrange's planetary equations
in the form
\begin{align}\label{lagrange}
\frac{da}{dt} &= \frac{2}{n a} \frac{\partial R}{\partial \sigma_{\text{b}}} +
      \left \langle \frac{da}{dt} \right \rangle_{\text{EF}} ~,
\notag \\
\frac{de}{dt} &= \frac{\alpha^{2}}{n a^{2} e}
      \frac{\partial R}{\partial \sigma_{\text{b}}} -
      \frac{\alpha}{n a^{2} e}
      \frac{\partial R}{\partial \tilde{\omega}} +
      \left \langle \frac{de}{dt} \right \rangle_{\text{EF}} ~,
\notag \\
\frac{d \sigma_{\text{b}}}{dt} + t \frac{dn}{dt} &= - \frac{2}{n a}
      \frac{\partial R}{\partial a} -
      \frac{\alpha^{2}}{n a^{2} e}
      \frac{\partial R}{\partial e} +
      \left \langle \frac{d \sigma_{\text{b}}}{dt} +
      t \frac{dn}{dt} \right \rangle_{\text{EF}} ~,
\notag \\
\frac{d \tilde{\omega}}{dt} &= \frac{\alpha}{n a^{2} e}
      \frac{\partial R}{\partial e} +
      \left \langle \frac{d \tilde{\omega}}{dt} \right \rangle_{\text{EF}} ~.
\end{align}
$\partial R / \partial a$ in Eqs. (\ref{lagrange}) is calculated in such a
way that $n$ is treated as a constant \citep{danby}. For this particular
case this is the only correct way and this also cancels the term
$\left \langle t ~dn / dt \right \rangle_{\text{G}}$ (where the subscript
$\text{G}$ denotes that the change is caused by gravitation only) in
the derivation resulting from $d \sigma_{1} / dt$ in Eqs. (\ref{transformed}).
During these operations the following relations have to be used:
\begin{align}\label{step}
\frac{\partial R}{\partial \sigma_{\text{b}}} &= - s
\frac{\partial R}{\partial \sigma_{0}} ~,
\notag \\
\frac{\partial R}{\partial \tilde{\omega}} &= - \frac{p + q}{q}
\frac{\partial R}{\partial \sigma_{0}} ~,
\notag \\
\frac{\partial R}{\partial \sigma_{0}} &= k \frac{\partial R}{\partial h} -
h \frac{\partial R}{\partial k} ~,
\notag \\
e \frac{\partial R}{\partial e} &= k \frac{\partial R}{\partial k} +
h \frac{\partial R}{\partial h} ~.
\end{align}
The relations obtained are
\begin{align}\label{relations}
\left \langle Q_{1} \right \rangle &= -
      \left \langle \frac{d \tilde{\omega}}{dt} \right \rangle_{\text{EF}} -
      \left \langle \frac{d \sigma_{\text{b}}}{dt} +
      t \frac{dn}{dt} \right \rangle_{\text{EF}} ~,
\notag \\
\left \langle Q_{2} \right \rangle &= -
      \left \langle \frac{d \tilde{\omega}}{dt} \right \rangle_{\text{EF}} ~,
\notag \\
\left \langle P_{1} \right \rangle &= \frac{n a}{2}
      \left \langle \frac{da}{dt} \right \rangle_{\text{EF}} ~,
\notag \\
\left \langle P_{2} \right \rangle &= - \frac{n a ( 1 - \alpha )}{2}
      \left \langle \frac{da}{dt} \right \rangle_{\text{EF}} -
      \frac{n a^{2} e}{\alpha}
      \left \langle \frac{de}{dt} \right \rangle_{\text{EF}} ~.
\end{align}
Eqs. (\ref{relations}) hold for arbitrary non-gravitational effects.

\section{The secular time derivative of the eccentricity of the orbit}
\label{sec:eccentricity}

The second equation of Eqs. (\ref{lagrange}) yields
\begin{align}\label{yield}
\frac{de}{dt} &= \frac{\alpha^{2}}{n a^{2} e}
      \frac{\partial R}{\partial \sigma_{\text{b}}} -
      \frac{\alpha}{n a^{2} e}
      \frac{\partial R}{\partial \tilde{\omega}} +
      \left \langle \frac{de}{dt} \right \rangle_{\text{EF}}
\notag \\
&= \frac{\alpha^{2}}{2 a e} \left ( \frac{da}{dt} -
      \left \langle \frac{da}{dt} \right \rangle_{\text{EF}} \right ) +
      \frac{\alpha}{n a^{2} e} \frac{p + q}{q}
      \frac{\partial R}{\partial \sigma_{0}} +
      \left \langle \frac{de}{dt} \right \rangle_{\text{EF}} ~,
\end{align}
where also the first equation of Eqs. (\ref{lagrange}) and
the second equation of Eqs. (\ref{step}) were used. If we use
the first equation of Eqs. (\ref{step}) and
the first equation of Eqs. (\ref{lagrange}) in Eq. (\ref{yield}),
then we obtain
\begin{align}\label{eccentricity}
\frac{de}{dt} &= \frac{\alpha^{2}}{2 a e} \left ( \frac{da}{dt} -
      \left \langle \frac{da}{dt} \right \rangle_{\text{EF}} \right ) -
      \frac{\alpha}{2 a e} \frac{p + q}{p} \left ( \frac{da}{dt} -
      \left \langle \frac{da}{dt} \right \rangle_{\text{EF}} \right ) +
      \left \langle \frac{de}{dt} \right \rangle_{\text{EF}}
\notag \\
&= \frac{\alpha}{2 a e} \left ( \alpha - \frac{p + q}{p} \right )
      \frac{da}{dt} + \frac{\alpha}{2 a e}
      \left ( \frac{p + q}{p} - \alpha \right )
      \left \langle \frac{da}{dt} \right \rangle_{\text{EF}} +
      \left \langle \frac{de}{dt} \right \rangle_{\text{EF}} ~.
\end{align}
This equation can be averaged over a resonant libration period
of the semimajor axis and the result is
\begin{equation}\label{libration}
\left \langle \frac{de}{dt} \right \rangle = \frac{\alpha}{2 a e}
\left ( \frac{p + q}{p} - \alpha \right )
\left \langle \frac{da}{dt} \right \rangle_{\text{EF}} +
\left \langle \frac{de}{dt} \right \rangle_{\text{EF}} ~,
\end{equation}
because
\begin{equation}\label{resonance}
\left \langle \frac{da}{dt} \right \rangle = 0 ~.
\end{equation}
Eq. (\ref{libration}) is consistent with the same result obtained
from an averaging of the time derivative of the Tisserand parameter over
the resonant libration period of the semimajor axis
(see \citealt{gomes97,CMDA,mmrflow}) and with the same result
obtained from a near-canonical approach \cite{gomes95,gomes97}.
We can rewrite Eq. (\ref{libration}) using the averaged
non-conservative terms in Eqs. (\ref{relations})
\begin{equation}\label{hybrid}
\left \langle \frac{de}{dt} \right \rangle = \frac{\alpha}{n a^{2} e}
\left ( \frac{1}{s} \left \langle P_{1} \right \rangle -
\left \langle P_{2} \right \rangle \right ) ~.
\end{equation}

\section{The effects that will be considered}
\label{sec:effects}

We take into account the electromagnetic radiation of the star, the radial
stellar wind, and the interstellar gas flow (IGF).

\subsection{Electromagnetic radiation}
\label{sec:elemag}

The acceleration of a homogeneous spherical dust particle caused by
the electromagnetic radiation of the star is given by the PR
effect \citep{poynting,robertson,icarus}. The acceleration
to first order in $v / c$ ($v$ is the speed of the dust particle with
respect to the star) has the form
\begin{equation}\label{PR}
\frac{d \vec{v}}{dt} = \beta \frac{\mu}{r^{2}}
\left [ \left ( 1 - \frac{\vec{v} \cdot \vec{e}_{\text{R}}}{c} \right )
\vec{e}_{\text{R}} - \frac{\vec{v}}{c} \right ] ~,
\end{equation}
where $r$ $=$ $\vert \vec{r} \vert$ is the distance from the star and
$\vec{e}_{\text{R}}$ $=$ $\vec{r} / r$ is the radial unit vector.

\subsection{Radial stellar wind}
\label{sec:wind}

The acceleration caused by the radial stellar wind is, to first
order in $v / c$, first order in $u / c$ ($u$ is the speed
of the stellar wind with respect to the star) and first order
in $v / u$ \citep[Eq.~37]{covsw}:
\begin{equation}\label{sw}
\frac{d \vec{v}}{dt} = \frac{\eta}{\bar{Q}'_{\text{pr}}}
\beta \frac{u}{c} \frac{\mu}{r^{2}} \left [
\left ( 1 - \frac{\vec{v} \cdot \vec{e}_{\text{R}}}{u} \right )
\vec{e}_{\text{R}} - \frac{\vec{v}}{u} \right ] ~.
\end{equation}
Here, $\eta$ is the ratio of the stellar wind energy to
the stellar electromagnetic radiation energy, both radiated per unit time
(to the given accuracy)
\begin{equation}\label{eta}
\eta = \frac{4 \pi r^{2} u}{L_{\star}}
\sum_{i = 1}^{N} n_{\text{sw}~i} m_{\text{sw}~i} c^{2} ~,
\end{equation}
where $m_{\text{sw}~i}$ and $n_{\text{sw}~i}$, $i$ $=$ 1 to $N$, are
the masses and concentrations of the stellar wind particles at a distance
$r$ from the star ($u$ $=$ 450 km/s and $\eta$ $=$ 0.38 for the Sun,
\citealt{covsw}).

\subsection{Interstellar gas flow}
\label{sec:flow}

The acceleration caused by the flow of neutral gas can be given in
the form \citep{baines}
\begin{equation}\label{acceleration}
\frac{d \vec{v}}{dt} = - \sum_{i = 1}^{N} c_{\text{D}i} \gamma_{i}
\vert \vec{v} - \vec{v}_{\text{F}} \vert
\left ( \vec{v} - \vec{v}_{\text{F}} \right ) ~.
\end{equation}
The sum in Eq. (\ref{acceleration}) runs over all particle species $i$.
$\vec{v}_{\text{F}}$ is the velocity of the IGF in the frame associated
with the star, $c_{\text{D}i}$ is the drag coefficient, and $\gamma_{i}$
is the collision parameter. The drag coefficient can be calculated from
\begin{align}\label{cd}
c_{\text{D}i}(s_{i}) = {} & \frac{1}{\sqrt{\pi}}
      \left ( \frac{1}{s_{i}} + \frac{1}{2 s_{i}^{3}} \right )
      \text{e}^{-s_{i}^{2}} +
      \left ( 1 + \frac{1}{s_{i}^{2}} - \frac{1}{4 s_{i}^{4}} \right )
      \text{erf}(s_{i})
\notag \\
& + \left ( 1 - \delta_{i} \right )
      \left ( \frac{T_{\text{d}}}{T_{i}} \right )^{1 / 2}
      \frac{\sqrt{\pi}}{3s_{i}} ~,
\end{align}
where erf$(s_{i})$ is the error function $\text{erf}(s_{i})$ $=$
$2 / \sqrt{\pi} \int_{0}^{s_{i}} \text{e}^{-t^{2}} dt$, $\delta_{i}$ is
the fraction of impinging particles specularly reflected at the surface
(for the rest of the particles, a diffuse reflection is assumed)
\citep{baines,gustafson}, $T_{\text{d}}$ is the temperature of the dust grain,
and $T_{i}$ is the temperature of the $i$th gas component.
$s_{i}$ is defined as a molecular speed ratio
\begin{equation}\label{s}
s_{i} = \sqrt{\frac{m_{i}}{2kT_{i}}} U ~.
\end{equation}
Here, $m_{i}$ is the mass of the neutral atom in the $i$th gas component,
$k$ is Boltzmann's constant, and
$U$ $=$ $\vert \vec{v} - \vec{v}_{\text{F}} \vert$
is the relative speed of the dust particle with respect to the gas.
For the collision parameter, we can write
\begin{equation}\label{cp}
\gamma_{i} = n_{i} \frac{m_{i}}{m} A ~,
\end{equation}
where $n_{i}$ is the concentration of the $i$th kind of interstellar
neutral atom, $A$ $=$ $\pi {R_{\text{d}}}^{2}$ is the geometric
cross section of the spherical dust grain, and $m$ is the grain mass.
Using the approximation that the speed of the IGF is much
greater than the speed of the dust grain in the frame
associated with the star ($\vert \vec{v} \vert$ $=$ $v$ $\ll$
$\vert \vec{v}_{\text{F}} \vert$ $=$ $v_{\text{F}}$),
an approximate expression for the relative speed is
\begin{equation}\label{cossent}
U = \vert \vec{v} - \vec{v}_{\text{F}} \vert =
\sqrt{v^{2} + v_{\text{F}}^{2} -
2 \vec{v} \cdot \vec{v}_{\text{F}}} \approx
v_{\text{F}} \left ( 1 - \frac{\vec{v} \cdot
\vec{v}_{\text{F}}}{v_{\text{F}}^{2}} \right ) ~.
\end{equation}
The same approximation can also be used in Eq. (\ref{cd}). The result is
\begin{align}\label{dragapprox}
c_{\text{D}i}(s_{i}) &\approx c_{\text{D}i}(s_{0i}) +
      \left ( \frac{dc_{\text{D}i}}{ds_{i}} \right )_{s_{i} = s_{0i}}
      ( s_{i} - s_{0i} )
\notag \\
&\equiv c_{\text{D}i}(s_{0i}) +
      \left ( \frac{dc_{\text{D}i}}{ds_{i}} \right )_{s_{i} = s_{0i}}
      \sqrt{\frac{m_{i}}{2kT_{i}}} ( U - v_{\text{F}} )
\notag \\
&\approx c_{0i} - k_{i}
      \frac{\vec{v} \cdot \vec{v}_{\text{F}}}{v_{\text{F}}} ~,
\end{align}
where
\begin{align}\label{short}
s_{0i} &\equiv \sqrt{\frac{m_{i}}{2kT_{i}}} v_{\text{F}} ~,
\notag \\
c_{0i} &\equiv c_{\text{D}i}(s_{0i}) ~,
\notag \\
k_{i} &\equiv \left ( \frac{dc_{\text{D}i}}{ds_{i}} \right )_{s_{i} = s_{0i}}
      \sqrt{\frac{m_{i}}{2kT_{i}}} ~.
\end{align}
Rewriting Eq. (\ref{acceleration}) using Eqs. (\ref{cossent}) and
(\ref{dragapprox}) yields
\begin{equation}\label{approxflow}
\frac{d \vec{v}}{dt} = - \sum_{i = 1}^{N} c_{0i} \gamma_{i} v_{\text{F}}^{2}
\biggl [ \frac{\vec{v}}{v_{\text{F}}} -
\frac{\vec{v}_{\text{F}}}{v_{\text{F}}} +
g_{i} \frac{\vec{v} \cdot \vec{v}_{\text{F}}}{v_{\text{F}}^{2}}
\frac{\vec{v}_{\text{F}}}{v_{\text{F}}} \biggr ] ~,
\end{equation}
where
\begin{equation}\label{gexpanded}
g_{i} = 1 + \frac{k_{i}}{c_{0i}} v_{\text{F}} =
\frac{1}{c_{0i}} \biggl [ \frac{1}{\sqrt{\pi}}
\left ( \frac{1}{s_{0i}} - \frac{3}{2s_{0i}^{3}} \right )
\text{e}^{-s_{0i}^{2}} +
\left ( 1 - \frac{1}{s_{0i}^{2}} + \frac{3}{4s_{0i}^{4}} \right )
\text{erf}(s_{0i}) \biggr ] ~.
\end{equation}
Eq. (\ref{approxflow}) was already used in \citet{dyncd} to derive
the secular time derivatives of the Keplerian orbital elements caused by
the IGF.

\section{The equation of motion}
\label{sec:EOM}

Since we want to study the motion of a dust particle in the frame of reference
associated with the star in the planar CR3BP, we must add to the sum of
Eqs. (\ref{PR}), (\ref{sw}), and (\ref{acceleration}), the corresponding
gravitational accelerations. If we assume that
$( \eta / \bar{Q}'_{\text{pr}} ) ( u / c )$ $\ll$ 1,
the equation of motion of the particle has the form
\begin{align}\label{EOM}
\frac{d \vec{v}}{dt} = {} & - \frac{\mu}{r^{2}}
      \left ( 1 - \beta \right ) \vec{e}_{\text{R}} -
      \frac{G_{0} M_{\text{P}}}{\vert \vec{r} - \vec{r}_{\text{P}} \vert^{3}}
      (\vec{r} - \vec{r}_{\text{P}}) -
      \frac{G_{0} M_{\text{P}}}{r_{\text{P}}^{3}}
      \vec{r}_{\text{P}}
\notag \\
& - \beta \frac{\mu}{r^{2}}
      \left ( 1 + \frac{\eta}{\bar{Q}'_{\text{pr}}} \right )
      \left ( \frac{\vec{v} \cdot \vec{e}_{\text{R}}}{c}
      \vec{e}_{\text{R}} + \frac{\vec{v}}{c} \right )
\notag \\
& - \sum_{i = 1}^{N} c_{\text{D}i} \gamma_{i}
      \vert \vec{v} - \vec{v}_{\text{F}} \vert
      \left ( \vec{v} - \vec{v}_{\text{F}} \right ) ~.
\end{align}
Eq. (\ref{EOM}) can be numerically solved in order to obtain the motion
of the dust particle.

\section{Averaged resonant equations for the non-gravitational
effects considered}
\label{sec:application}

Because we assume that the electromagnetic radiation and radial stellar
wind are independent of the direction from the star (spherical symmetry),
the secular variations of any given particle orbit are independent
of the spatial orientation of the orbit for the accelerations caused by
the PR effect and the radial stellar wind. Therefore, for the PR effect
and the radial stellar wind, the theory presented in \citet{FM1} can
be used. However, the secular variations of the particle orbit for
the acceleration caused by an IGF depend on the orientation
of the orbit with respect to the IGF velocity vector and in this
case the theory developed in Section \ref{sec:average} can be used. The last
term in Eq. (\ref{EOM}) does not allow directly determining the secular
variations of the particle orbit caused by the IGF in a finite
analytical form. Hence, in analytical calculations, we will use
Eq. (\ref{approxflow}) instead of Eq. (\ref{acceleration}). Thus,
the non-conservative part of the acceleration in Eq. (\ref{grad}) is
\begin{align}\label{Y}
\vec{Y} = {} & - \beta \frac{\mu}{r^{2}}
      \left ( 1 + \frac{\eta}{\bar{Q}'_{\text{pr}}} \right )
      \left ( \frac{\vec{v} \cdot \vec{e}_{\text{R}}}{c}
      \vec{e}_{\text{R}} + \frac{\vec{v}}{c} \right )
\notag \\
& - \sum_{i = 1}^{N} c_{0i} \gamma_{i} v_{\text{F}}^{2}
      \biggl [ \frac{\vec{v}}{v_{\text{F}}} -
      \frac{\vec{v}_{\text{F}}}{v_{\text{F}}} +
      g_{i} \frac{\vec{v} \cdot \vec{v}_{\text{F}}}{v_{\text{F}}^{2}}
      \frac{\vec{v}_{\text{F}}}{v_{\text{F}}} \biggr ] ~.
\end{align}
We can calculate the average values of $Q_{i}$ and $P_{i}$ in Eqs. (\ref{PQ})
using the assumption that the temperature, concentration, and
velocity of all gas components in the IGF are constant. The results
for the averaged non-conservative terms in Eqs. (\ref{transformed}) are
\begin{align}\label{PQ_averaged}
\left \langle Q_{1} \right \rangle &= - \sum_{i = 1}^{N}
      \frac{3 c_{0i} \gamma_{i}
      v_{\text{F}} a e S}{L} + \left ( 1 - \alpha \right )
      \left \langle Q_{2} \right \rangle ~,
\notag \\
\left \langle Q_{2} \right \rangle &=
      \sum_{i = 1}^{N} \frac{c_{0i} \gamma_{i} v_{\text{F}} a \alpha S}{2 L}
      \left \{ \frac{3}{e} - \frac{\sigma g_{i} I}{v_{\text{F}}}
      \left [ \frac{2 \alpha^{2}}{\left ( 1 + \alpha \right )^{2}} -
      1 \right ] \right \} ~,
\notag \\
\left \langle P_{1} \right \rangle &= - \frac{\beta \mu n}{2 c \alpha^{3}}
      \left ( 1 + \frac{\eta}{\bar{Q}'_{\text{pr}}} \right )
      \left ( 3 e^{2} + 2 \right ) - \sum_{i = 1}^{N} c_{0i} \gamma_{i}
      v_{\text{F}}^{2} \sigma a \alpha \left [ 1 + \frac{g_{i}
      \left ( S^{2} + \alpha I^{2} \right )}
      {v_{\text{F}}^{2} \left ( 1 + \alpha \right )} \right ] ~,
\notag \\
\left \langle P_{2} \right \rangle &= -
      \left \langle P_{1} \right \rangle - \frac{\beta \mu n}{c}
      \left ( 1 + \frac{\eta}{\bar{Q}'_{\text{pr}}} \right ) -
      \sum_{i = 1}^{N} c_{0i} \gamma_{i} v_{\text{F}}^{2} a
      \left [ \frac{3 e I}{2 v_{\text{F}}} + \sigma \alpha^{2}
      \left ( 1 + \frac{g_{i}}{2} \right ) \right ] ~,
\end{align}
where
\begin{equation}\label{sigma}
\sigma = \frac{1}{v_{\text{F}}}
\sqrt{\frac{\mu ( 1 - \beta )}{a ( 1 - e^{2} )}} ~,
\end{equation}
\begin{align}\label{SI}
S &= v_{\text{F}1} \cos \tilde{\omega} + v_{\text{F}2} \sin \tilde{\omega} ~,
\notag \\
I &= - v_{\text{F}1} \sin \tilde{\omega} + v_{\text{F}2} \cos \tilde{\omega} ~,
\end{align}
with the Cartesian coordinates of the interstellar gas flow velocity
vector denoted by $v_{\text{F}k}$. An equivalent way of obtaining
Eqs. (\ref{PQ_averaged}) is to use the following identities in
Eqs. (\ref{relations}):
\begin{align}\label{secular}
\left \langle \frac{da}{dt} \right \rangle_{\text{EF}} = {} & -
      \frac{\beta \mu}{c a \alpha^{3}}
      \left ( 1 + \frac{\eta}{\bar{Q}'_{\text{pr}}} \right )
      \left ( 2 + 3 e^{2} \right )
\notag \\
& - \sum_{i = 1}^{N} \frac{2 c_{0i} \gamma_{i} v_{\text{F}}^{2}
      \sigma a^{2} \alpha}{L}
      \left [ 1 + \frac{g_{i} \left ( S^{2} + \alpha I^{2} \right )}
      {v_{\text{F}}^{2} \left ( 1 + \alpha \right )} \right ] ~,
\notag \\
\left \langle \frac{de}{dt} \right \rangle_{\text{EF}} = {} & -
      \frac{\beta \mu}{2 c a^{2} \alpha}
      \left ( 1 + \frac{\eta}{\bar{Q}'_{\text{pr}}} \right ) 5 e
\notag \\
& + \sum_{i = 1}^{N} \frac{c_{0i} \gamma_{i} v_{\text{F}} a \alpha}{2 L}
      \left [ 3 I + \frac{\sigma g_{i} \alpha^{2} \left ( 1 - \alpha \right )
      \left ( I^{2} - S^{2} \right )}
      {v_{\text{F}} e \left ( 1 + \alpha \right )} \right ] ~,
\notag \\
\left \langle \frac{d \tilde{\omega}}{dt} \right \rangle_{\text{EF}} = {} &
      \sum_{i = 1}^{N} \frac{c_{0i} \gamma_{i} v_{\text{F}} a \alpha S}{2 L}
      \left \{ - \frac{3}{e} +
      \frac{\sigma g_{i} I}{v_{\text{F}}}
      \left [ \frac{2 \alpha^{2}}{\left ( 1 + \alpha \right )^{2}} -
      1 \right ] \right \} ~,
\notag \\
\left \langle \frac{d \sigma_{\text{b}}}{dt} +
t \frac{dn}{dt} \right \rangle_{\text{EF}} = {} &
      \sum_{i = 1}^{N} \frac{c_{0i} \gamma_{i} v_{\text{F}} a S}{2 L}
      \left \{ \frac{3 \left ( 1 + e^{2} \right )}{e} -
      \frac{\sigma g_{i} \alpha^{2} I}{v_{\text{F}}}
      \left [ \frac{2 \alpha^{2}}{\left ( 1 + \alpha \right )^{2}} -
      1 \right ] \right \} ~.
\end{align}
The first three equations in Eqs. (\ref{secular}) were already obtained
in \citet{dyncd} in a slightly different, but equivalent, form. In
the first equation we can see that the semimajor axis is always a decreasing
function of time. Hence, the particles can migrate into MMRs
from larger values of semimajor axes. The evolution
of particle's Keplerian orbital elements under the action
of the considered non-gravitational effects in the planar case
(when the interstellar gas flow velocity lies in the orbital plane
of the particle) without the gravitational influence of a planet is
characterized by non-monotonic variations in the orbital eccentricity.
The eccentricity can commonly reach values larger than 0.5 and then can
again decrease to values close to zero \citep{flow,fgf}. At the values
of eccentricity close to zero the dust particle can be captured into MMRs
more efficiently. A detailed analysis of capture probability at the values
of eccentricity larger than 0.5 is beyond the scope of this paper.
However, we will return to this open problem in Section \ref{sec:orbits}.

The time derivative of the orbit eccentricity averaged over the resonant
libration period of the semimajor axis, obtained from
Eq. (\ref{libration}) or Eq. (\ref{hybrid}), for the considered
effects, is
\begin{align}\label{planar}
\left \langle \frac{de}{dt} \right \rangle = {} &
      \frac{\beta \mu \alpha}{c a^{2} e}
      \left ( 1 + \frac{\eta}{\bar{Q}'_{\text{pr}}} \right )
      \left ( 1 - \frac{2 + 3 e^{2}}{2 \alpha^{3}} \frac{p + q}{p} \right )
\notag \\
& + \sum_{i = 1}^{N} \frac{c_{0i} \gamma_{i} v_{\text{F}}^{2} a \alpha}{L}
      \Biggl \{ \frac{3}{2} \frac{I}{v_{\text{F}}} +
      \frac{\sigma \alpha^{2}}{e} \left ( 1 + \frac{g_{i}}{2} \right )
\notag \\
& - \frac{\sigma \alpha}{e} \Biggl [ 1 +
      \frac{g_{i} \left ( S^{2} + \alpha I^{2} \right )}
      {v_{\text{F}}^{2} \left ( 1 + \alpha \right )} \Biggr ]
      \frac{p + q}{p} \Biggr \} ~.
\end{align}

\section{Comparison of the analytical and numerical results}
\label{sec:comparison}

Interstellar gas also penetrates into the heliosphere and can affect
the motion of dust particles in the outer parts of the Solar system.
The IGF in the Solar system arrives from the direction
$\lambda_{\text{ecl}}$ $=$ 254.7$^{\circ}$ (heliocentric ecliptic longitude)
and $\beta_{\text{ecl}}$ $=$ 5.2$^{\circ}$ (heliocentric ecliptic latitude;
\citealt{lallement}). The IGF contains mainly hydrogen and helium atoms. Some
of the atoms are ionized. The ionized hydrogen in the IGF can
acquire electrons from interstellar H$^{\circ}$ in the outer heliosheath
during the passage into the heliopause \citep{frisch,alouani}.
Therefore, there are two populations of neutral interstellar hydrogen inside
the heliopause. The primary population comprises the original neutral
hydrogen atoms of the IGF which penetrated into the heliopause. The secondary
population is created by charge exchange in the outer heliosheath
(between the bow shock and the heliopause). Neutral atoms from both
the primary and the secondary population penetrate freely inside
the heliopause. The solar wind concentration decreases as $r^{-2}$ inside
the heliopause. Thus, the efficiency of charge exchange is reduced
dramatically as compared with the efficiency in front of the heliopause,
and the two populations flow through the inner heliosheath practically without
any change \citep{frisch}. We neglect the influences of solar gravity and solar
radiation pressure on the motion of the neutral hydrogen inside the heliopause,
which is important only within a few AUs from the Sun. For the same reason, we
also neglect any losses of neutral hydrogen due to ionization by the solar wind
and the solar extreme ultraviolet radiation inside the heliopause. Neutral
interstellar helium flows freely through the outer heliosphere
(with only $\le$ 2\% filtration through charge exchange with H$^{+}$)
and inside of the Earth's orbit is ionized by photoionization and
electron ionization \citep{frisch}. We assume that the temperatures,
concentrations, and velocities of both populations of the neutral hydrogen
and the neutral helium can be approximated by constant values. To see that
this approximation is usable, we refer the reader to \citet{alouani}.

\begin{table}[t]
\centering
\begin{tabular}{c c c c c}
\hline
\hline
Fig. & $a_{\text{in}}$ & $e_{\text{in}}$ & $\tilde{\omega}_{\text{in}}$ &
$f_{\text{in}}$\\
$ $ & [ AU ] & [ - ] & [ $^{\circ}$ ] & [ $^{\circ}$ ]\\
\hline
\ref{fig:stabilization} & $a_{2 / 1}$ $+$ 0.001 & 0.6 & 295 & 0\\
\ref{fig:oscillation} & $a_{2 / 1}$ $+$ 0.001 & 0.5 & 0 & 0\\
\ref{fig:top} & $a_{2 / 1}$ $+$ 0.001 & 0.5 & 90 & 0\\
\hline
\end{tabular}
\caption{The initial conditions of the dust particle for the numerical
solutions of the equation of motion of which the results were depicted in
Figs. \ref{fig:stabilization}, \ref{fig:oscillation} and \ref{fig:top}.}
\label{T1}
\end{table}

The velocity vector of the IGF in the Solar system does not lie
exactly in Neptune's orbital plane. The angle between the direction
of the velocity vector of the interstellar gas and Neptune's orbital plane is
3.7$^{\circ}$. The obtained orbital evolutions of the dust particles
under the action of the PR effect, the radial solar wind, and the IGF are
almost indistinguishable from the coplanar case due to the small value
of this angle \citep{flow}. This can be understood using
the secular time derivatives of the orbital elements for
an arbitrary orientation of the orbit caused by the considered
effects. The secular time derivatives are shown in \citet{dyncd}.
The secular time derivative of the inclination caused by the PR effect
and the radial solar wind is zero, and the secular time derivative
of the inclination caused by the IGF is proportional to the component
of the velocity vector of the interstellar gas that is normal
to the particle's orbital plane (in our case, approximately
the orbital plane of Neptune). Because the normal
component of the interstellar gas velocity vector is small,
the inclination of the orbit can be approximated by a constant
value (in our case, a value close to zero). The fact that
the velocity vector of the IGF does not lie exactly in
the orbital plane of Neptune has also a positive effect.
The velocity and concentration of the IGF are not significantly affected by
the Sun, because no part of Neptune's orbit is
in the ``shadow'' cast by the Sun in the IGF. The ``shadow''
goes beneath Neptune's orbit.

We have adopted the following concentrations and temperatures
for the various components in the IGF.
$n_{1}$ $=$ 0.059 cm$^{-3}$ and $T_{1}$ $=$ 6100 K
for the primary population of neutral hydrogen \citep{frisch},
$n_{2}$ $=$ 0.059 cm$^{-3}$ and $T_{2}$ $=$ 16500 K
for the secondary population of neutral hydrogen \citep{frisch}, and
$n_{3}$ $=$ 0.015 cm$^{-3}$ and $T_{3}$ $=$ 6300 K for the neutral helium
\citep{lallement}. The assumed interstellar gas speed is equal for all
components and identical to the speed of the neutral helium entering
the Solar system, $v_{\text{F}}$ $=$ 26.3 km s$^{-1}$ \citep{lallement}.
The IGF velocity vector was rotated into Neptune's orbital plane
around an axis lying in Neptune's orbital plane and perpendicular
to the velocity vector of the interstellar gas, in order to ensure
the validity of the planar case in our numerical investigations.

\begin{figure}[t]
\begin{center}
\includegraphics[width=0.8\textwidth]{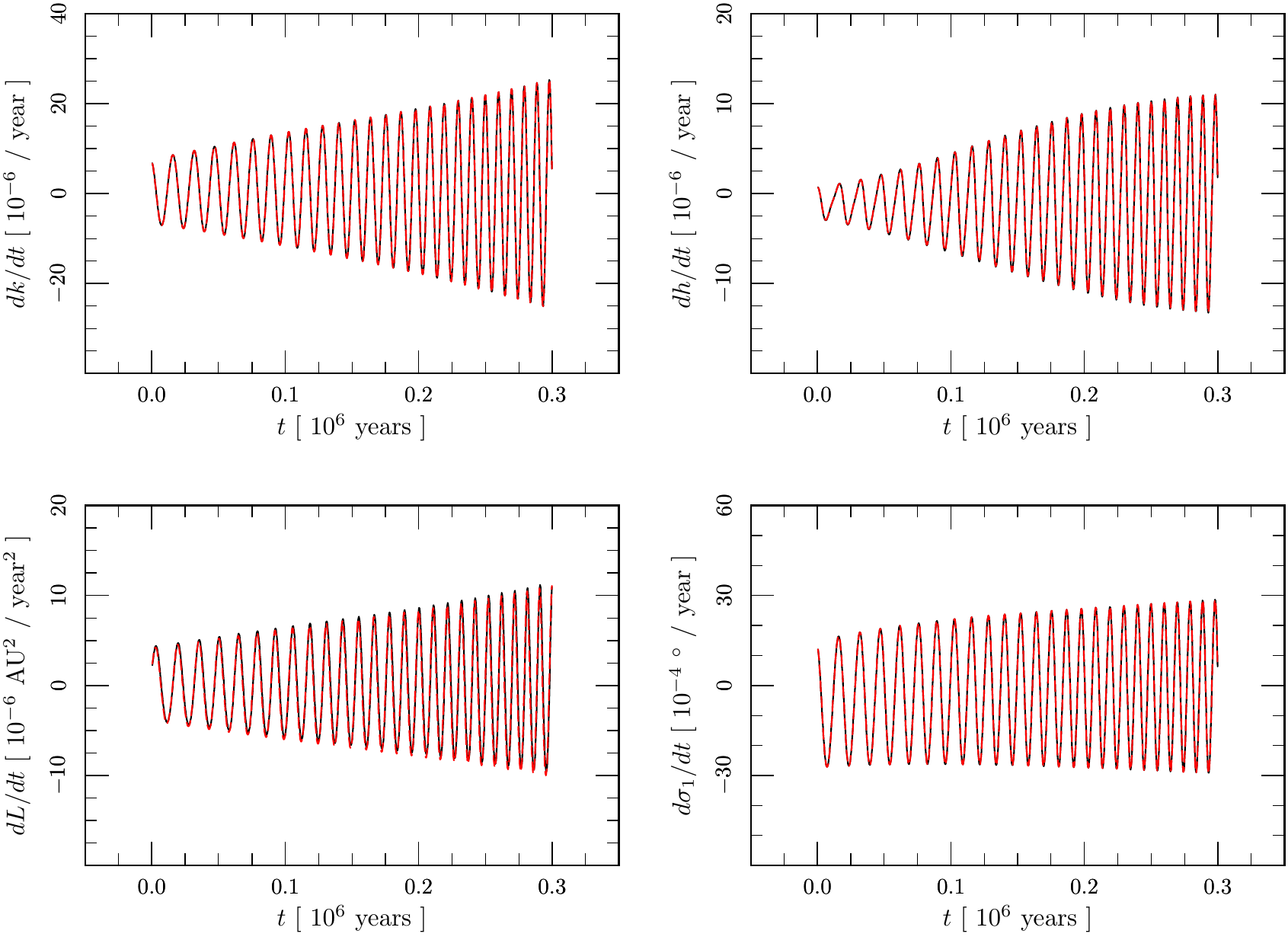}
\end{center}
\caption{Comparison of the time derivatives of $k$, $h$, $L$ and $\sigma_{1}$,
averaged over the synodic period, obtained from the numerical solution
of the equation of motion (solid black line) and from the analytical theory
(dashed red line) for a dust particle with $R_{\text{d}}$ $=$ 2 $\mu$m,
$\varrho$ $=$ 1 g/cm$^{3}$, and $\bar{Q}'_{\text{pr}}$ $=$ 1 captured
in an exterior mean motion orbital 2/1 resonance with Neptune under
the action of the PR effect, the radial solar wind, and the IGF.}
\label{fig:derivatives}
\end{figure}

To compare the analytical and numerical results, we considered
a dust particle with $R_{\text{d}}$ $=$ 2 $\mu$m,
$\varrho$ $=$ 1 g/cm$^{3}$, and $\bar{Q}'_{\text{pr}}$ $=$ 1.
We neglected the Lorentz force, which is only important for
submicrometer particles \citep{dohnanyi,leinert,dermott}. The interval
between collisions is on the order of 10$^{7}$ years for a particle
with $R_{\text{d}}$ $=$ 2 $\mu$m beyond the orbit of Neptune \citep{flow}.
We assumed that the atoms are specularly reflected at the surface
of the dust grain ($\delta_{i}$ $=$ 1 in Eq. \ref{cd}), and we used
the approximation that the drag coefficients are constant ($c_{\text{D}i}$
$=$ $c_{\text{D}i}(s_{0i})$ and $g_{i}$ $=$ 1). The approximation
that the drag coefficients are constant is usable if
$\vert \vec{v} \vert$ $\ll$ $v_{\text{F}}$
holds during an orbit (for a comparison of the evolutions, see
Fig. 5 in \citealt{dyncd}). The time derivatives of the parameters $k$, $h$,
$L$, and $\sigma_{1}$, averaged over a synodic period, obtained from
a numerical solution of Eq. (\ref{EOM}) (solid black line) and from
analytical relations given by Eqs. (\ref{transformed}) (dashed red
line) are compared in Fig. \ref{fig:derivatives}. The planet was initially
located on the positive $x$-axis. The initial semimajor axis of the dust
particle for the numerical solution of equation of motion is computed
from the relation $a_{\text{in}}$ $=$ $a_{n_{\text{P}} / n}$
$+$ $\triangle$, where $a_{n_{\text{P}} / n}$ $=$ $a_{\text{P}}$
$( 1 - \beta )^{1/3}$ $( n_{\text{P}} / n )^{2/3}$ (with
the assumption that $M_{\text{P}}$ $\ll$ $M_{\star}$) and $\triangle$ is
the shift from the exact resonant semimajor axis. As the initial conditions for
the dust particle, we used $a_{\text{in}}$ $=$ $a_{2 / 1}$
$+$ 0.001 AU, $e_{\text{in}}$ $=$ 0.2, $\tilde{\omega}_{\text{in}}$ $=$
285$^{\circ}$. The initial true anomaly of the dust particle was
$f_{\text{in}}$ $=$ 0$^{\circ}$. On the right-hand sides
of Eqs. (\ref{transformed}) we used the numerically
calculated values of $k$, $h$, $L$, $a$, $e$
and $\tilde{\omega}$ averaged over the synodic period.
The terms $\partial R / \partial a$, $\partial R / \partial e$, and
$\partial R / \partial \tilde{\omega}$ averaged over
the synodic period were calculated from their
definition given by Eq. (\ref{DF}). $\partial R / \partial k$ and
$\partial R / \partial h$ in Eqs. (\ref{transformed}) were calculated
from Eqs. (\ref{step}) using $\partial R / \partial e$ and
$\partial R / \partial \sigma_{0}$. As can be seen in
Fig. \ref{fig:derivatives}, the numerical and analytical results are
in excellent agreement.

\begin{figure}[t]
\begin{center}
\includegraphics[width=0.8\textwidth]{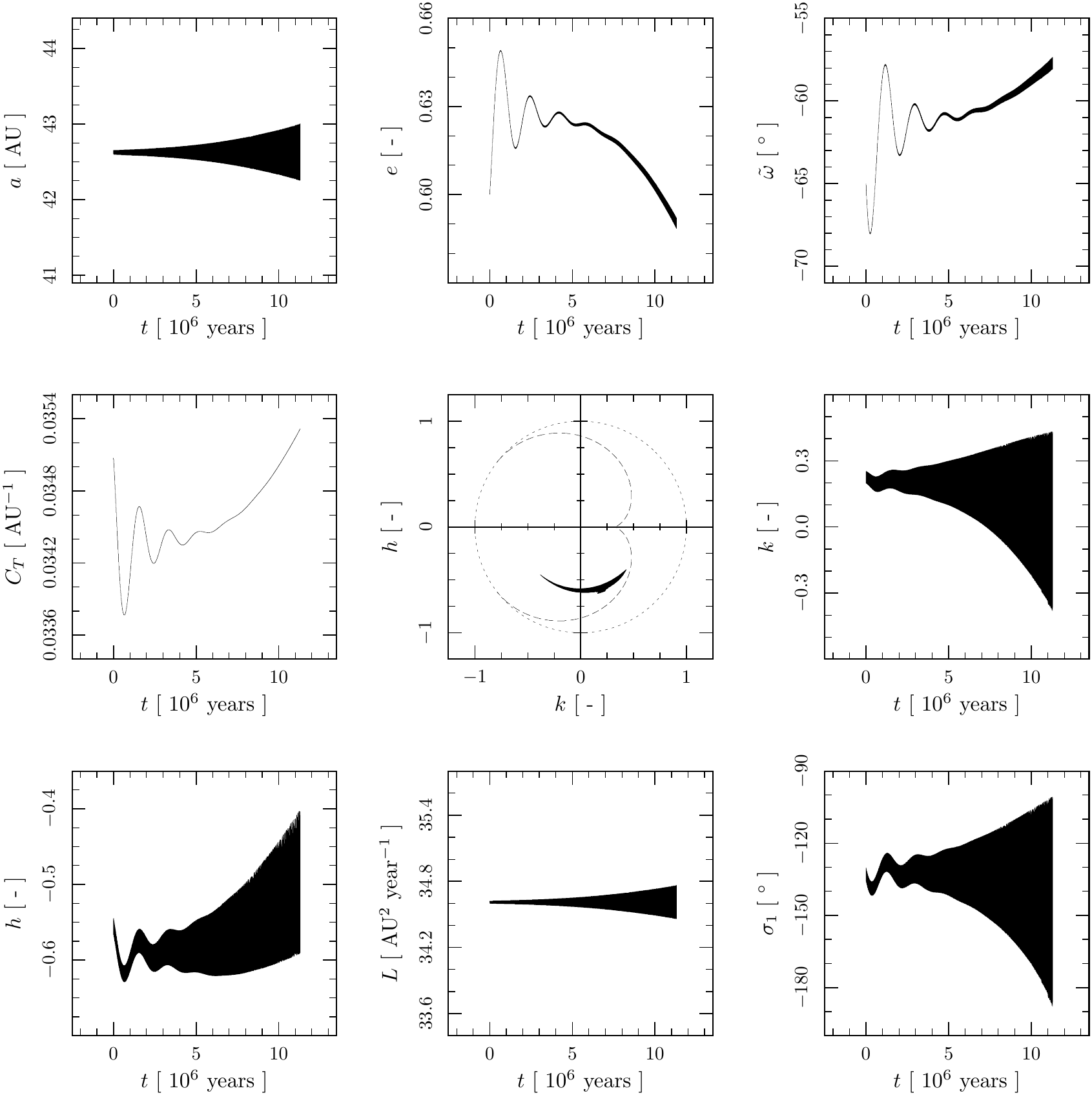}
\end{center}
\caption{The evolutions of the parameters described in the text
for a dust particle with $R_{\text{d}}$ $=$ 2 $\mu$m,
$\varrho$ $=$ 1 g/cm$^{3}$, and $\bar{Q}'_{\text{pr}}$ $=$ 1
captured in an exterior mean motion orbital 2/1 resonance with
Neptune under the action of the PR effect, the radial solar wind,
and the IGF. In the $kh$ plane, the resonant libration occurs
around the bottom libration center of the conservative problem.
The evolutions of the eccentricity and longitude of perihelion
during libration around this center have damped oscillations
approaching some relatively slowly changing values.}
\label{fig:stabilization}
\end{figure}

\begin{figure}[t]
\begin{center}
\includegraphics[width=0.8\textwidth]{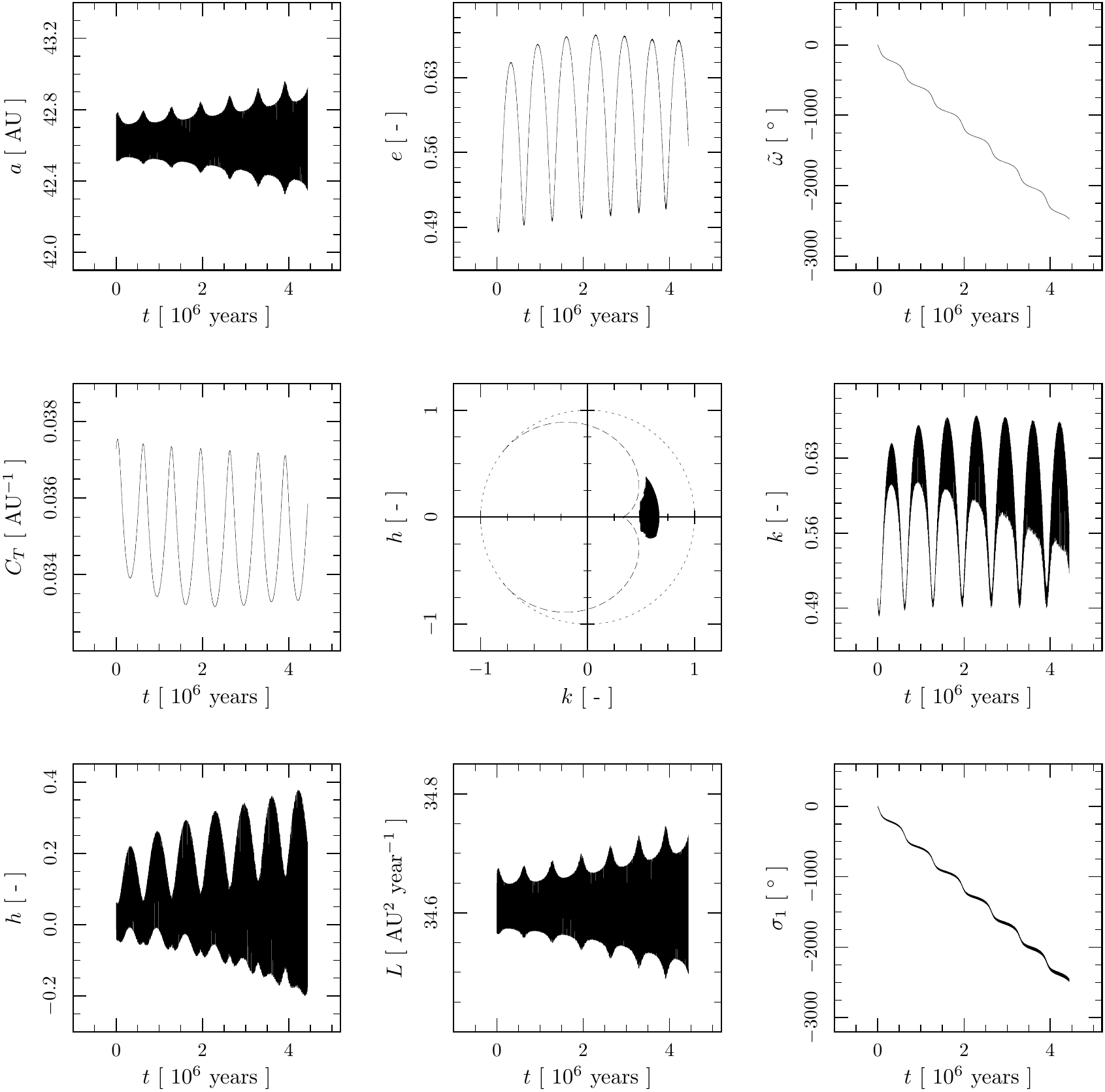}
\end{center}
\caption{The same dust particle, resonance, and non-gravitational
effects as in Fig. \ref{fig:stabilization}, but in this case the resonance
libration occurs around the libration center of the conservative problem
which is located on the right side of the origin in the $kh$ plane.
The orbital evolutions with libration around this center have oscillations
in the eccentricity and a fast monotonic advance of the longitude
of perihelion.}
\label{fig:oscillation}
\end{figure}

\begin{figure}[t]
\begin{center}
\includegraphics[width=0.8\textwidth]{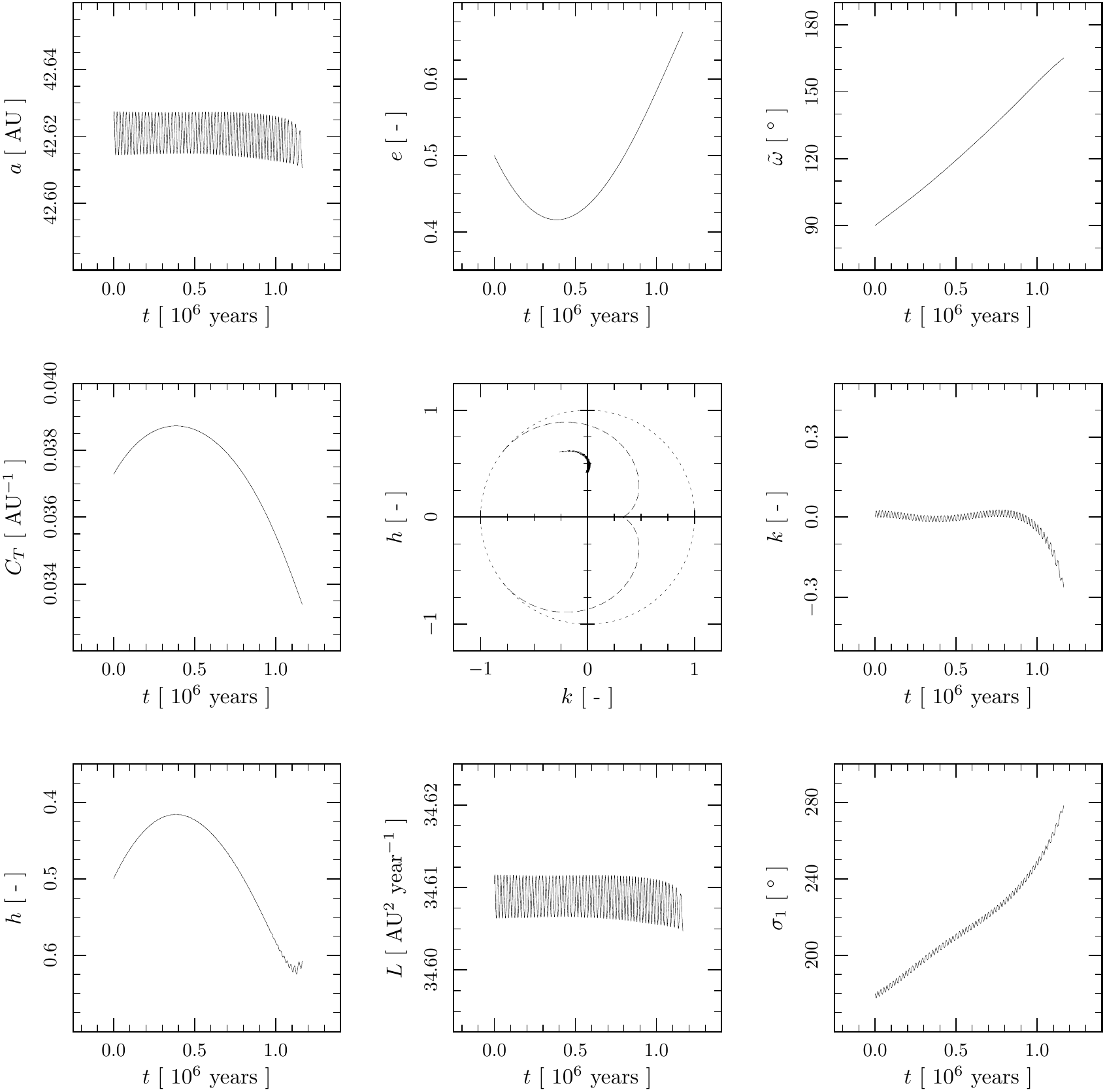}
\end{center}
\caption{The same dust particle, resonance, and non-gravitational
effects as in Figs. \ref{fig:stabilization} and \ref{fig:oscillation}, but
the particle is captured by the resonant libration around the top libration
center of the conservative problem in the $kh$ plane.}
\label{fig:top}
\end{figure}

\section{Types of orbits}
\label{sec:orbits}

The evolution in the $kh$ plane is found to be an ideal tool for
determining the types of orbits. The main property of this plane
is its division into sets by a line on which collisions
of the planet and the particle take place. The collision line,
for the exterior mean motion orbital 2/1 resonance, is depicted
in Figs. \ref{fig:stabilization}, \ref{fig:oscillation}, and \ref{fig:top}
by a dashed line. For this resonance, the phase space is practically
divided into two sets. We will refer to the first set as the region
below the collision curve (low-eccentricity region) and the second set
as the region beyond the collision curve. The evolutions in the region
below the collision curve are depicted in Figs. \ref{fig:stabilization} and
\ref{fig:top}. One evolution from the region beyond the collision
curve is shown in Fig. \ref{fig:oscillation}. We used
the same parameters for the dust particle and
the non-gravitational effects as in Section \ref{sec:comparison}.
The evolutions depicted are, from left to right and from top to bottom:
$a$ (semimajor axis), $e$ (eccentricity), $\tilde{\omega}$
(longitude of perihelion), $C_{\text{T}}$ $=$ $(1 - \beta) / (2 a) +
\sqrt{( 1 - \beta ) a ( 1 - e^{2} ) / a_{\text{P}}^{3}}$
(Tisserand parameter), $kh$ point, $k$ $=$ $e \cos \sigma_{0}$,
$h$ $=$ $e \sin \sigma_{0}$, $L$ $=$ $\sqrt{\mu ( 1 - \beta ) a}$, and
$\sigma_{1}$ $=$ $( p + q ) \lambda_{\text{P}} / q$ $-$ $p \lambda / q$.
All parameters are averaged over the synodic period.
We must note that the evolution duration 11.3 $\times$ 10$^{6}$ years
(Fig. \ref{fig:stabilization}) requires the size
of an interstellar gas cloud to be 303.9 pc in the direction
of the interstellar gas velocity vector (a constant velocity
with magnitude 26.3 km s$^{-1}$ is assumed) and such a situation
cannot always occur in the real Galactic environment. Evolutions with
such long capture times should be used only for theoretical purposes.
A shorter part of the evolution is still valid. The planet was initially
located on the positive $x$-axis. The initial conditions of the dust
particle are in Table \ref{T1}. We used numerical integrations
of the equation of motion to decide if for the values of eccentricity
close to 0.5 (used as the initial eccentricities in
Figs. \ref{fig:stabilization}--\ref{fig:top}) it is possible to capture
the dust particle into an exterior mean motion orbital 2/1 resonance with
Neptune under the action of the considered non-gravitational effects or not.
We found that the particle can be captured into the 2/1 commensurability
even for larger eccentricities. From a sample of 144 dust particles with
the same physical properties as in Figs. \ref{fig:stabilization}--\ref{fig:top},
the number of particles captured into the MMR was 7
for $e_{\text{in}}$ $=$ 0.5 and 33 for $e_{\text{in}}$ $=$ 0.01
all captures occurred during 10$^{5}$ years after beginning of the numerical
solution. The particles were initially 1 AU from the exact 2/1 commensurability
($\triangle$ $=$ 1 AU) and with uniformly distributed initial
longitudes of perihelion and initial true anomalies.

An evolution in the region beyond the collision curve correspond to
a libration around one of the libration centers of the conservative problem,
which is located on the right of the origin on the $h$-axis.
The orbital evolutions in the region beyond the collision curve have
a fast monotonic advance of the perihelion and oscillations
of the eccentricity (Fig. \ref{fig:oscillation}).

The orbital evolutions in the region below the collision
curve can be divided into two groups, corresponding
to two libration centers of the conservative problem. One evolution
with a libration around the center located below the $h$-axis
(bottom center) is depicted in Fig. \ref{fig:stabilization}
and one evolution with a libration around the center located above
the $h$-axis (top center) in Fig. \ref{fig:top}. The evolutions
of eccentricity and longitude of perihelion during
the libration around the bottom center have damped oscillations
approaching some ``constant'' values. We found that these values
are not exactly constant, but they are relatively slowly changing
in comparison with the evolution before this ``stabilization'' (as can
be seen in Fig. \ref{fig:stabilization}).

\begin{figure}[t]
\begin{center}
\includegraphics[width=0.5\textwidth]{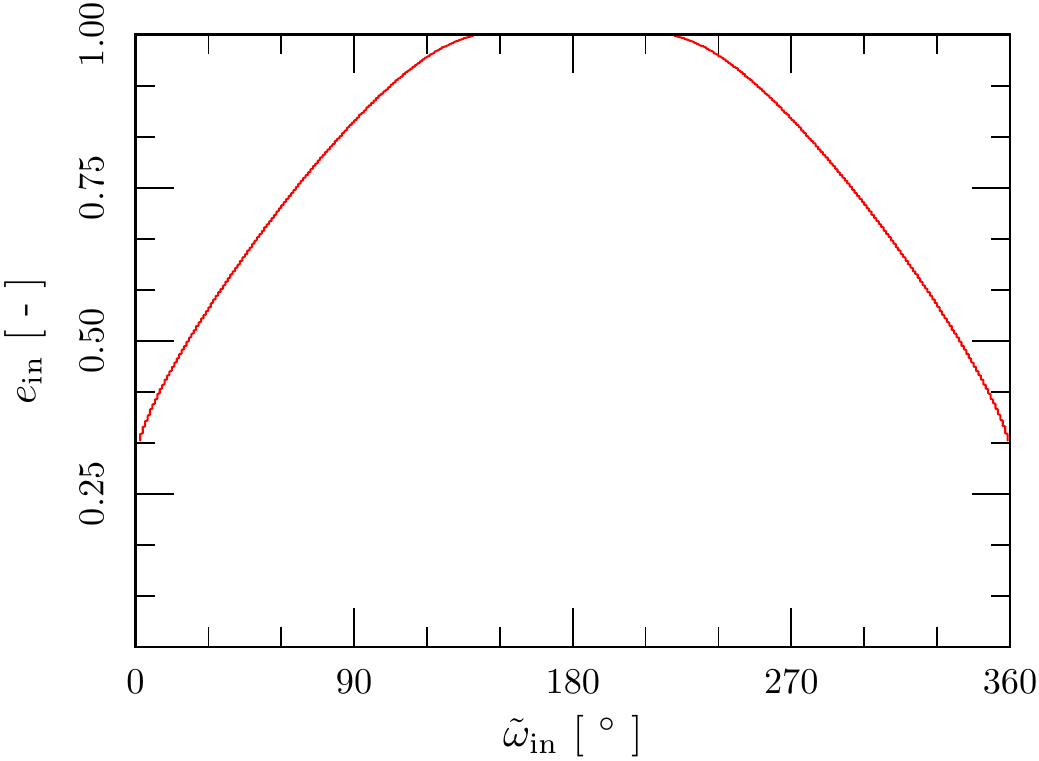}
\end{center}
\caption{In the case when a planet in a circular orbit is initially
located on the positive $x$-axis and a dust particle with
$R_{\text{d}}$ $=$ 2 $\mu$m, $\varrho$ $=$ 1 g/cm$^{3}$, and
$\bar{Q}'_{\text{pr}}$ $=$ 1 is initially in the perihelion,
the depicted curve corresponds to the initial longitudes
of perihelion and eccentricities of the particle leading
to collisions of the planet with the particle for the exact
exterior mean motion orbital 2/1 resonance.}
\label{fig:mmrflow}
\end{figure}

To compare the maximal capture times for evolutions corresponding
to the various libration centers of the conservative problem, we divided
the initial eccentricity, longitude of perihelion, true anomaly
of the particle, and true anomaly of the planet uniformly within the allowed
intervals for the exterior mean motion orbital 2/1 resonance. The number
of captures and the maximal capture times which we found for the various
libration centers were different. The orbital evolutions with
libration around the bottom center have larger maximal capture times
than the evolutions librating around the top center. This is
different from the conservative problem, where both libration centers
are symmetric with respect to the $h$-axis \cite{beauge} and also
have equal maximal capture times. We found that a horseshoe-like
libration between the top and the bottom libration centers in
the region below the collision curve is also possible
with smaller maximal capture times.

The two groups with the largest maximal capture times (libration
in the region beyond the collision curve and libration around the bottom
center) were already identified in \citet{mmrflow}. For a planet initially
located on the positive $x$-axis and a particle with initial conditions
$a_{\text{in}}$ $=$ $a_{2 / 1}$, $e_{\text{in}}$ arbitrary,
$\tilde{\omega}_{\text{in}}$ arbitrary, and $f_{\text{in}}$ $=$
0$^{\circ}$, we obtain, in the $\tilde{\omega}_{\text{in}} e_{\text{in}}$
plane, the collision line depicted in Fig. \ref{fig:mmrflow}.
This curve divides the evolutions in \citet{mmrflow} into two groups
(in the present paper, we refer to these two groups as
the region below the collision curve and
the region beyond the collision curve).

\section{Conditions for stationary points}
\label{sec:conditions}

Equations (\ref{transformed}) enable searching for stationary points
determined by
\begin{equation}\label{fixed}
\frac{dk}{dt} = \frac{dh}{dt} = \frac{dL}{dt} = 0 ~.
\end{equation}
Eqs. (\ref{fixed}) are conditions for the existence of a fixed point
in the subset created by the averaged values of $k$, $h$ and $L$.
This should not be understood in the sense that if the conditions
hold, then the point will be stationary during an arbitrary
time interval. The stationarity is obtained only in the synodic period.
Substitution of Eqs. (\ref{fixed}) in Eqs. (\ref{transformed}) leads to
the conditions
\begin{align}\label{conditions}
\left \langle P_{1} \right \rangle -
s \left \langle P_{2} \right \rangle &= 0 ~,
\notag \\
e \frac{\partial R}{\partial e} = k \frac{\partial R}{\partial k} +
      h \frac{\partial R}{\partial h} &=
      \frac{L e^{2}}{\alpha} \left ( \frac{d \sigma_{1}}{dt} +
      \left \langle Q_{2} \right \rangle \right ) ~,
\notag \\
\frac{\partial R}{\partial \sigma_{0}} = k \frac{\partial R}{\partial h} -
h \frac{\partial R}{\partial k} &= \left \langle P_{2} \right \rangle ~.
\end{align}
The first equation in Eqs. (\ref{conditions}) is equivalent
to the condition $\langle de / dt \rangle$ $=$ 0
(Eq. \ref{hybrid}).

Equations (\ref{conditions}) enable also stationary points
with $d \sigma_{1} / dt$ $\neq$ 0, in general.
Now we show that such stationary points are not possible
under the action of the considered non-gravitational effects.
The second equation of Eqs. (\ref{conditions}) can be rewritten using
the second equation of Eqs. (\ref{relations}) and the last equation
of Eqs. (\ref{lagrange}) as follows:
\begin{equation}\label{omegas}
e \frac{\partial R}{\partial e} =
\frac{L e^{2}}{\alpha} \left ( \frac{d \sigma_{1}}{dt} -
\frac{d \tilde{\omega}}{dt} + \frac{\alpha}{L e}
\frac{\partial R}{\partial e} \right ) ~.
\end{equation}
This equation is equivalent to the condition
\begin{equation}\label{sigmas}
\frac{d \sigma_{0}}{dt} = \frac{d \sigma_{1}}{dt} -
\frac{d \tilde{\omega}}{dt} = 0 ~.
\end{equation}
Therefore, if $d \sigma_{1} / dt$ $\neq$ 0, then the longitude
of pericenter must change. Since the time derivative
of the eccentricity is zero for a stationary point, we can write, for
the first of Eqs. (\ref{conditions}) (see Eqs. \ref{PQ_averaged}),
\begin{equation}\label{constant}
B(\tilde{\omega}) = \left \langle P_{1} \right \rangle -
s \left \langle P_{2} \right \rangle =
\alpha_{1} S^{2} + \alpha_{2} I^{2} + \alpha_{3} I + \alpha_{4} = 0 ~,
\end{equation}
where $\alpha_{l}$ for $l$ $=$ 1, .., 4 are some constants.
$B(\tilde{\omega})$ is constant for all $\tilde{\omega}$ and
its derivative with respect to time must be zero. We get
\begin{equation}\label{dB}
\frac{dB}{dt} = \frac{dB}{d \tilde{\omega}} \frac{d \tilde{\omega}}{dt} =
\left ( 2 \alpha_1 I - 2 \alpha_2 I - \alpha_{3} \right ) S
\frac{d \tilde{\omega}}{dt} = 0 ~.
\end{equation}
Because $\alpha_1$ $\neq$ $\alpha_2$ and $\alpha_3$ $\neq$ 0,
for all existing stationary points there must hold
\begin{equation}\label{result}
\frac{d \tilde{\omega}}{dt} = \frac{d \sigma_{1}}{dt} = 0 ~.
\end{equation}
We conclude that the time derivative of the longitude of pericenter
must be zero for all stationary points. Identities Eqs. (\ref{fixed}),
Eqs. (\ref{sigmas}) and Eqs. (\ref{result}) imply that equations
\begin{equation}\label{orbit}
\frac{da}{dt} = \frac{de}{dt} = \frac{d \tilde{\omega}}{dt} = 0
\end{equation}
hold for the stationary points. Therefore, we can call them stationary
orbits. If we use Eqs. (\ref{fixed}) and Eqs. (\ref{result}) in
Eqs. (\ref{transformed}), we obtain for the considered non-gravitational
effects the following conditions for the stationary orbits:
\begin{align}\label{system}
\left \langle P_{1} \right \rangle -
s \left \langle P_{2} \right \rangle &= 0 ~,
\notag \\
\frac{\partial R}{\partial e} &= \frac{L e}{\alpha}
      \left \langle Q_{2} \right \rangle ~,
\notag \\
\frac{\partial R}{\partial \sigma_{0}} &= \left \langle P_{2} \right \rangle ~,
\notag \\
\frac{2 s L}{\mu ( 1 - \beta )} \frac{\partial R}{\partial a} &= -
      n_{\text{P}} \frac{p + q}{q} + n s -
      s \left \langle Q_{1} \right \rangle +
      \left ( 1 - \alpha \right ) s \left \langle Q_{2} \right \rangle ~.
\end{align}
Solutions of this system of equations represent points in
a five-dimensional space with $a_{\text{in}}$, $e_{\text{in}}$,
$\tilde{\omega}_{\text{in}}$, $f_{\text{in}}$ and $f_{\text{P in}}$ on
the axes. Here, $f_{\text{in}}$ and $f_{\text{P in}}$ are the initial true
anomalies of the particle and the planet, respectively.

\section{Searching for stationary orbits in an exterior mean motion
orbital 2/1 resonance with Neptune}
\label{sec:Neptune}

Equations (\ref{system}) can be used to search for orbits stationary
over a synodic period. The approach can be divided into the special case
of exact resonance and the more general case of non-exact resonance.

\subsection{Exact resonance}
\label{sec:exact}

In this subsection, the symbols used denote the non-averaged
quantities. In the first part of this subsection, we determine
some properties of the searched stationary orbits in the exact
resonance. We assume that the system does not change during
the synodic period. For the time derivative of an arbitrary quantity
$\Psi$ averaged over the synodic period $T_{\text{S}}$, we can write
\begin{equation}\label{Psi}
\frac{1}{T_{\text{S}}} \int_{0}^{T_{\text{S}}} \frac{d \Psi}{dt} dt =
\frac{\Psi(T_{\text{S}}) - \Psi(0)}{T_{\text{S}}} ~.
\end{equation}
A zero average value of the time derivative means that the quantity
$\Psi$ returns to its value at the time zero after the synodic period.
Eqs. (\ref{orbit}) imply that the semimajor axis, the eccentricity, and
the longitude of pericenter all return to their values at the time zero
after the synodic period. In other words, the evolution leading to
the same orbit after one synodic period satisfies these identities.
In our case, the non-averaged resonant angular variable is
\begin{equation}\label{setup}
\sigma_{0} = \frac{p + q}{q} \lambda_{\text{P}} -
      s \lambda - \tilde{\omega} ~,
\end{equation}
with the mean longitude
\begin{equation}\label{lambda}
\lambda = M + \tilde{\omega} = n t + \sigma_{\text{b}} + \tilde{\omega} ~.
\end{equation}
Using this notation we obtain for the time derivative of $\sigma_{0}$
\begin{equation}\label{dsigma0dt}
\frac{d \sigma_{0}}{dt} = \frac{p + q}{q} n_{\text{P}} - s n -
s \left ( \frac{d \sigma_{\text{b}}}{dt} + t \frac{dn}{dt} +
\frac{d \tilde{\omega}}{dt} \right ) -
\frac{d \tilde{\omega}}{dt} ~.
\end{equation}
Eq. (\ref{dsigma0dt}) can be averaged over the synodic period. If we use
Eqs. (\ref{sigmas}) and Eqs. (\ref{result}) for a stationary orbit
in the result of averaging, we obtain
\begin{equation}\label{passage}
\frac{p + q}{q} n_{\text{P}} - \frac{s}{T_{\text{S}}}
\int_{0}^{T_{\text{S}}} n ~dt -
\frac{s}{T_{\text{S}}} \int_{0}^{T_{\text{S}}}
\left ( \frac{d \sigma_{\text{b}}}{dt} +
t \frac{dn}{dt} \right ) dt = 0 ~.
\end{equation}
This equation is equivalent to the last equation in
Eqs. (\ref{system}), as can be easily showed using
the third equation in Eqs. (\ref{lagrange}), the first
and the second equation in Eqs. (\ref{relations}),
and the second equation in Eqs. (\ref{system}).
An exact resonance with
\begin{equation}\label{ratio}
n_{\text{P}} (p + q)  = \frac{p}{T_{\text{S}}} \int_{0}^{T_{\text{S}}} n ~dt
\end{equation}
reduces Eq. (\ref{passage}) to the form
\begin{equation}\label{exact}
\frac{1}{T_{\text{S}}} \int_{0}^{T_{\text{S}}}
\left ( \frac{d \sigma_{\text{b}}}{dt} + t \frac{dn}{dt} \right ) dt = 0 ~.
\end{equation}
We can calculate the time derivative of the non-averaged mean longitude
in Eqs. (\ref{lambda}):
\begin{equation}\label{dlambdadt}
\frac{d \lambda}{dt} = n + \frac{d \sigma_{\text{b}}}{dt} + t \frac{dn}{dt} +
\frac{d \tilde{\omega}}{dt} ~.
\end{equation}
This equation can also be averaged over the synodic period.
If we also use Eqs. (\ref{result}) and Eqs. (\ref{exact}), then we obtain
\begin{equation}\label{average}
\frac{1}{T_{\text{S}}} \int_{0}^{T_{\text{S}}} \frac{d \lambda}{dt} dt =
\frac{1}{T_{\text{S}}} \int_{0}^{T_{\text{S}}} n ~dt ~.
\end{equation}
Eq. (\ref{average}) can be rewritten using Eq. (\ref{ratio}) and
Eq. (\ref{Psi})
\begin{equation}\label{delta}
\frac{\lambda(T_{\text{S}}) - \lambda(0)}{T_{\text{S}}} =
\frac{p+q}{p} n_{\text{P}} ~.
\end{equation}
Eq. (\ref{result}) gives that $\tilde{\omega}(T_{\text{S}})$ $=$
$\tilde{\omega}(0)$. Hence, Eq. (\ref{delta}) is
\begin{equation}\label{anomaly}
M(T_{\text{S}}) - M(0) = \frac{p+q}{p} n_{\text{P}} T_{\text{S}} ~.
\end{equation}
In the next step, we determine $T_{\text{S}}$.
Implicit differentiation of the angular variable $\sigma_{2}$ defined
in Eqs. (\ref{regular}) yields
\begin{equation}\label{differentiation}
d \sigma_{2} = \frac{1}{q} \left ( n ~dt + t ~dn +
d \sigma_{\text{b}} + d \tilde{\omega} - n_{\text{P}} ~dt \right ) ~.
\end{equation}
The last equation can be integrated over the synodic period
\begin{equation}\label{integration}
2 \pi = \frac{1}{q} \left [ \int_{0}^{T_{\text{S}}} n ~dt +
\int_{0}^{T_{\text{S}}} \left ( \frac{d \sigma_{\text{b}}}{dt} +
t ~\frac{dn}{dt} \right ) dt +
\int_{0}^{T_{\text{S}}} \frac{d \tilde{\omega}}{dt} dt -
n_{\text{P}} ~T_{\text{S}} \right ] ~,
\end{equation}
where we have used the fact that the change of $\sigma_{2}$ is $2 \pi$ after
the synodic period. Substituting Eqs. (\ref{result}) and Eq. (\ref{exact})
into Eq. (\ref{integration}) yields
\begin{equation}\label{one}
T_{\text{S}} = \frac{1}{n_{\text{P}}} \left ( \int_{0}^{T_{\text{S}}} n ~dt -
2 \pi q \right ) ~.
\end{equation}
The integral in Eq. (\ref{one}) can be expressed using
Eq. (\ref{ratio}) in order to obtain the synodic period
\begin{equation}\label{TS}
T_{\text{S}} = \frac{2 \pi p}{n_{\text{P}}} ~.
\end{equation}
The substitution of Eq. (\ref{TS}) in Eq. (\ref{anomaly})
gives the following results for the stationary
orbit in the exact resonance after one synodic period
\begin{equation}\label{whole}
M(T_{\text{S}}) - M(0) = \left ( p + q \right ) 2 \pi ~.
\end{equation}
After one synodic period, the orbit is the same. The position
of the particle on the orbit can be determined using
the mean anomaly. Because the difference between the mean anomalies
after the synodic period is an integer times $2 \pi$, the positions
at time zero and at time $T_{\text{S}}$ are the same.
The motion which satisfies Eqs. (\ref{orbit}) and
Eq. (\ref{whole}) is periodic after the synodic period. The second
implication holds trivially. The motion periodic after the synodic
period satisfies these identities. Hence, Eqs. (\ref{orbit})
and Eq. (\ref{whole}) imply that the orbit will be stable during
an arbitrary time interval (during which the conditions hold).
The evolution in the synodic period will be repeated periodically.
Such a stable orbit is characterized by a zero amplitude of resonant
libration. We must note that in reality other small forces can
destroy the exact periodicity.

\begin{figure}[t]
\begin{center}
\includegraphics[width=0.5\textwidth]{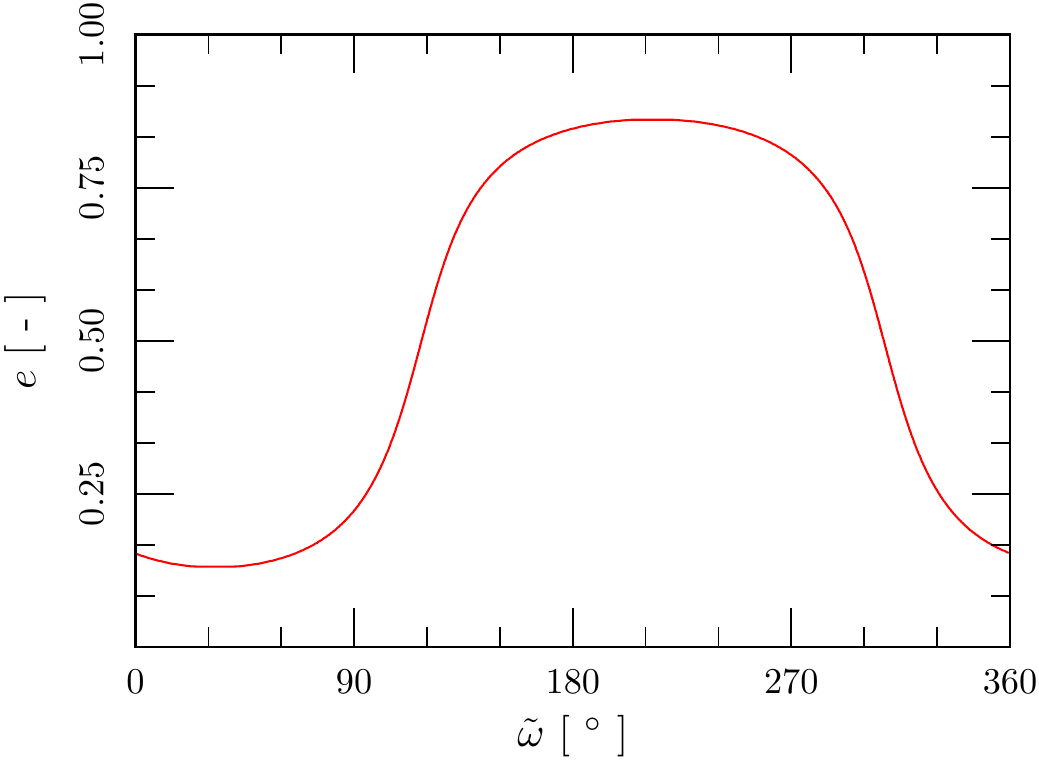}
\end{center}
\caption{The longitudes of perihelion and the eccentricities
for which the first equation of Eqs. \ref{system} holds for a dust
particle with $R_{\text{d}}$ $=$ 2 $\mu$m, $\varrho$ $=$ 1 g/cm$^{3}$, and
$\bar{Q}'_{\text{pr}}$ $=$ 1 captured in an exterior mean motion orbital 2/1
resonance with Neptune under the action of the PR effect, the radial solar
wind, and the IGF.}
\label{fig:zero}
\end{figure}

\begin{figure}[t]
\begin{center}
\includegraphics[width=0.8\textwidth]{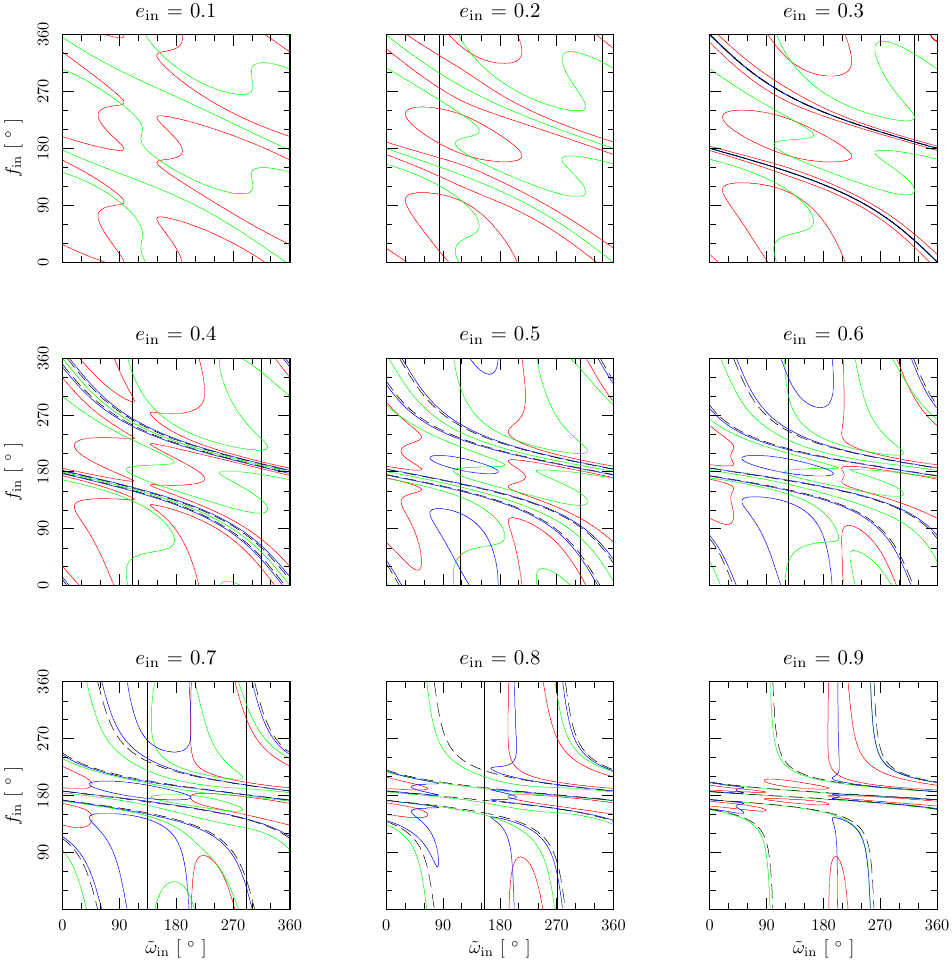}
\end{center}
\caption{Sections of three-dimensional phase space determined by
solutions of the last three equations in Eqs. (\ref{system}) for
a dust particle with $R_{\text{d}}$ $=$ 2 $\mu$m,
$\varrho$ $=$ 1 g/cm$^{3}$, and $\bar{Q}'_{\text{pr}}$ $=$ 1
captured in the exact exterior mean motion orbital 2/1 resonance
with Neptune under the action of the PR effect, the radial solar wind,
and the IGF. Vertical lines in a given section correspond
to the longitude of perihelion for which the first equation
of Eqs. (\ref{system}) holds. Red, green, and blue lines
correspond to solutions of the second, third, and fourth equation
in Eqs. (\ref{system}), respectively. Collisions of the planet with
the particle occur for points on the dashed lines.}
\label{fig:cube}
\end{figure}

\begin{figure}[t]
\begin{center}
\includegraphics[width=0.5\textwidth]{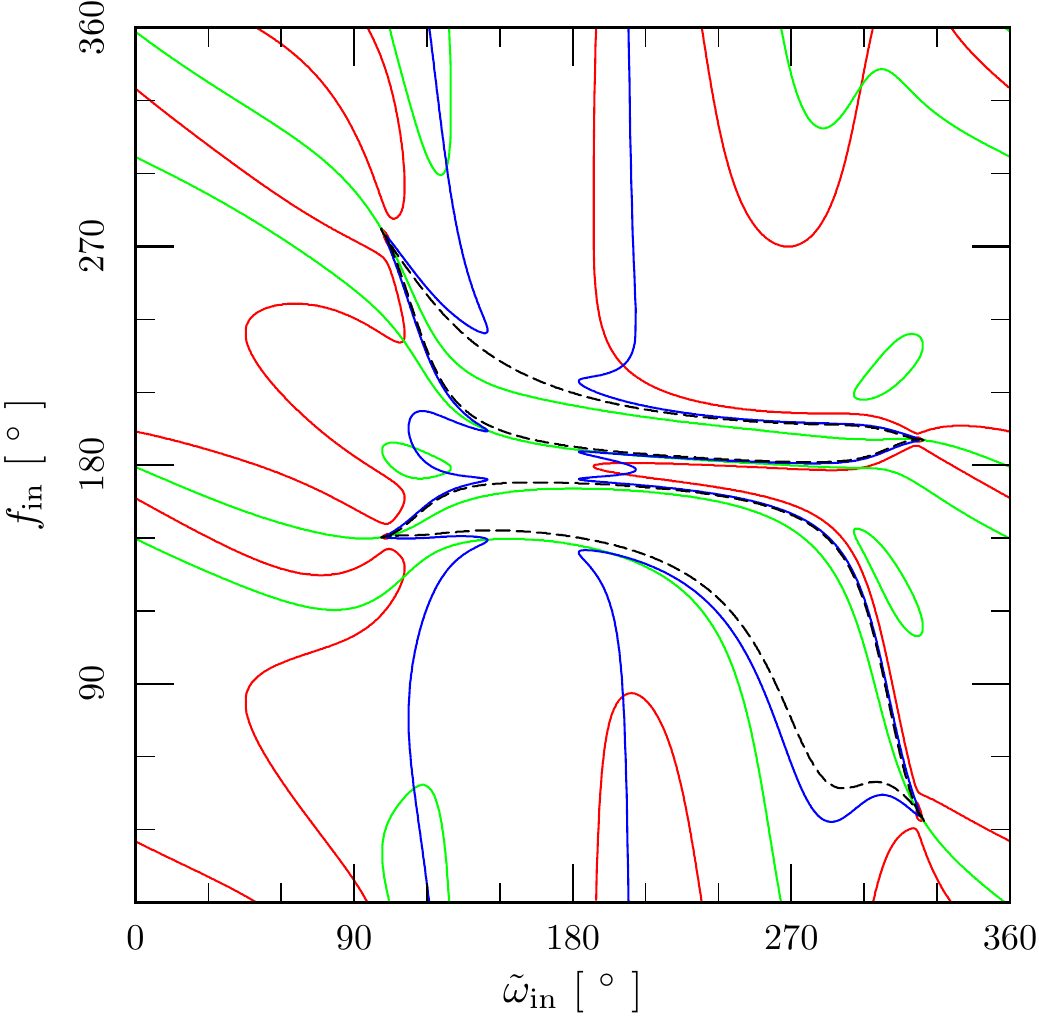}
\end{center}
\caption{Red, green, and blue lines correspond to solutions
of the second, third, and fourth equation in Eqs. (\ref{system}) with
substituted solutions from the first equation in Eqs. (\ref{system}),
respectively. Collisions of the planet with the particle occur for points
on the dashed lines.}
\label{fig:substitution}
\end{figure}

The previous part of this subsection described the properties
of the searched stable orbits in the exact resonance.
As was already mentioned, the phase space of the solutions
of the system of equations given by Eqs. (\ref{system})
has five dimensions. For an exact resonance, the number of dimensions can
be reduced. These reductions will be implemented in the numerical
calculation of the stable orbits in the exact resonance. If we consider
the exact resonance, then $a_{\text{in}}$ is determined by
the condition $a_{\text{in}}$ $=$ $a_{\text{P}}$ $( 1 - \beta )^{1/3}$
$\left [ M_{\star} / ( M_{\star} + M_{\text{P}}) \right ]^{1/3}$
$[ p / ( p + q ) ]^{2/3}$. Due to the exact resonance, a further
reduction in the number of dimensions is possible. In the exact resonance
we have
\begin{equation}\label{reduction}
\frac{1}{T_{\text{S}}} \int_{0}^{T_{\text{S}}}
\frac{\partial R}{\partial e} d t = \frac{1}{T_{\text{S}}}
\left ( \int_{T_{\text{A}}}^{T_{\text{S}}}
\frac{\partial R}{\partial e} d t +
\int_{T_{\text{S}}}^{T_{\text{S}} + T_{\text{A}}}
\frac{\partial R}{\partial e} d t \right ) ~,
\end{equation}
here $T_{\text{A}}$ denotes the time at which the planet is in a fixed
initial position. The validity of Eq. (\ref{reduction}) is caused by the fact
that for a stable orbit in the exact resonance, the relative positions
of the planet and the dust particle repeat periodically, and therefore
\begin{equation}\label{periodicity}
\frac{1}{T_{\text{S}}} \int_{0}^{T_{\text{A}}}
\frac{\partial R}{\partial e} d t = \frac{1}{T_{\text{S}}}
\int_{T_{\text{S}}}^{T_{\text{S}} + T_{\text{A}}}
\frac{\partial R}{\partial e} d t ~.
\end{equation}
Similar equations hold for the averaged $\partial R / \partial \sigma_{0}$
and $\partial R / \partial a$. In other words, in the exact resonance,
for all initial positions of the planet and all initial positions
of the dust particle, we can find the same value of the terms
$\partial R / \partial e$, $\partial R / \partial \sigma_{0}$, and
$\partial R / \partial a$ averaged over the synodic period for
a fixed initial position of the planet and calculated initial positions
of the dust particle. Hence, the solutions of the last three
equations in the system of equations given by Eqs. (\ref{system})
for the exact resonance represent points in a three-dimensional
space with $e_{\text{in}}$, $\tilde{\omega}_{\text{in}}$, and
$f_{\text{in}}$ on the axes.

\begin{figure}[t]
\begin{center}
\includegraphics[width=0.8\textwidth]{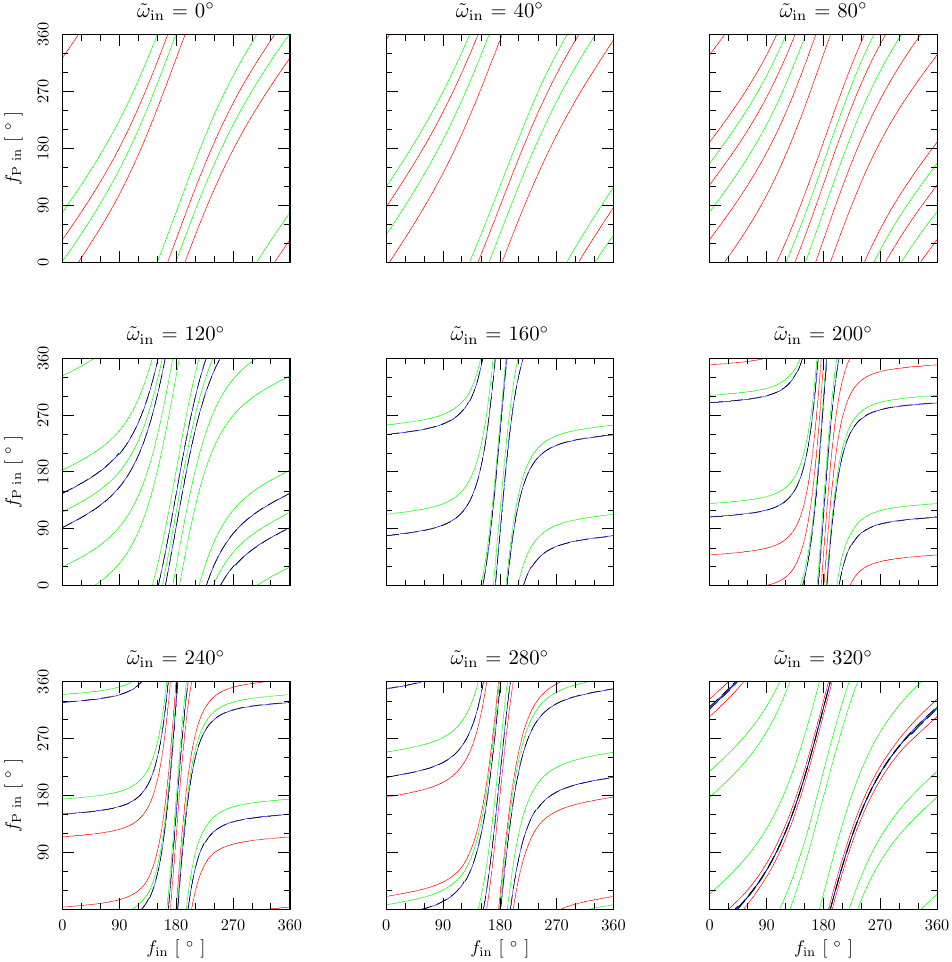}
\end{center}
\caption{Solutions of the second (red line), third (green line) and fourth
equation (blue line) with substituted solutions of the first equation
in the system given by Eqs. (\ref{system}) for a dust
particle with $R_{\text{d}}$ $=$ 2 $\mu$m, $\varrho$ $=$ 1 g/cm$^{3}$, and
$\bar{Q}'_{\text{pr}}$ $=$ 1 captured in a non-exact exterior mean motion
orbital 2/1 resonance with Neptune under the action of the PR effect,
the radial solar wind, and the IGF. The shift from the exact resonance
is $\triangle$ $=$ 0.075 AU. Collisions of the planet with the particle
occur for points on the dashed lines.}
\label{fig:shift-p}
\end{figure}

\begin{figure}[t]
\begin{center}
\includegraphics[width=0.8\textwidth]{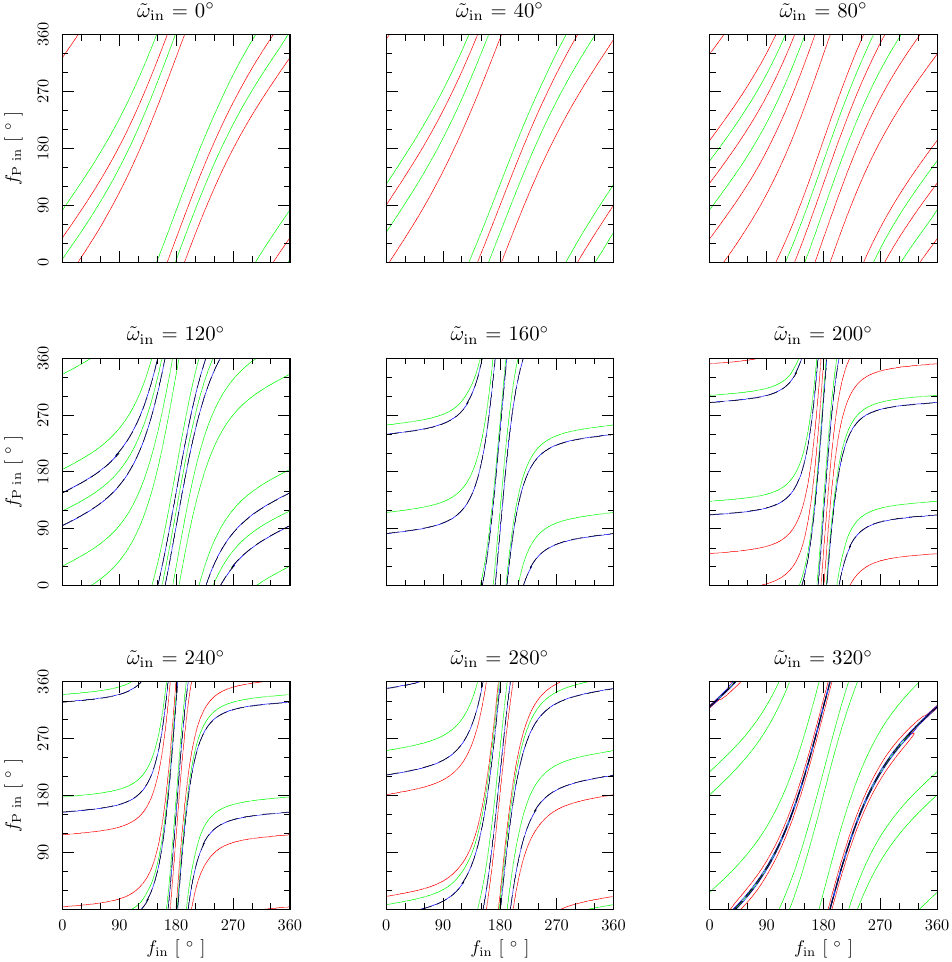}
\end{center}
\caption{The same as in Fig. \ref{fig:shift-p}, but in this case
the shift from the exact resonance is $\triangle$ $=$ -0.075 AU.}
\label{fig:shift-n}
\end{figure}

\begin{figure}[t]
\begin{center}
\includegraphics[width=0.8\textwidth]{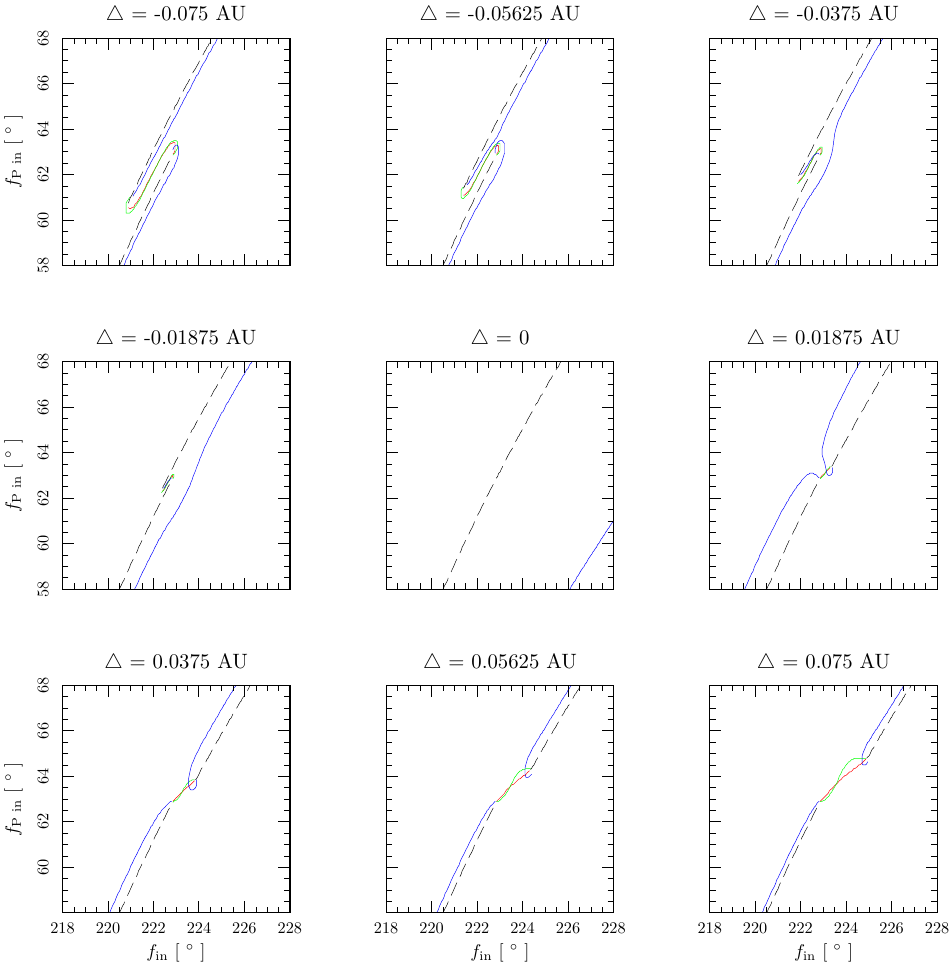}
\end{center}
\caption{A region in the $f_{\text{in}} f_{\text{P in}}$ plane
obtained with $\tilde{\omega}_{\text{in}}$ $=$ 200$^{\circ}$.
The region is centered on the initial conditions that directly lead
to a collision for $\triangle$ = 0 AU. The same situation as in
Figs. \ref{fig:shift-p} and \ref{fig:shift-n} is solved, but in this
case various values of the shift in an arithmetic sequence are used.}
\label{fig:zoom}
\end{figure}

We have searched for the stable orbits for a particle captured in the exact
exterior mean motion orbital 2/1 resonance with Neptune under the action
of the PR effect, the radial solar wind, and the IGF. The particle and
non-gravitational effects were described by the same parameters as in
Section \ref{sec:comparison}. The relation depicted in Fig. \ref{fig:zero}
represents pairs of $\tilde{\omega}$ and $e$ for which the first equation
of Eqs. (\ref{system}) holds. Any stable solution must lie on this
curve. Also the secular time derivative of the eccentricity
averaged over a libration period is zero for the points on the curve
(Section \ref{sec:eccentricity}). Sections of the three-dimensional
phase space of the solutions of the three remaining equations in the system
of equations given by Eqs. (\ref{system}) for the exact resonance
are shown for nine different initial eccentricities in Fig. \ref{fig:cube}.
The fixed initial position of the planet was on the positive $x$-axis.
The red, green, and blue points correspond to solutions of the second,
third, and fourth equation in Eqs. (\ref{system}), respectively. Any point
on the dashed curves in Fig. \ref{fig:cube} leads to a collision
of the planet with the particle. The perihelion distance is larger than
the radius of the planet's orbit for $e_{\text{in}}$ $\apprle$ 0.2945.
Therefore, the dashed lines are not on the sections with initial eccentricity
equal to 0.1 and 0.2. Since the planet's initial position is fixed,
for every orbit with $e_{\text{in}}$ $\apprge$ 0.2945, there exist some
true anomalies of the particle which lead to a collision.
The vertical lines in a given section are on the positions given
by the longitudes of perihelion for which the first equation
of Eqs. (\ref{system}) holds (see also Fig. \ref{fig:zero}).
Fig. \ref{fig:cube} allows determining the stable orbits in
the exact resonance under the action of the PR effect, the radial stellar
wind, and the IGF. As can be seen in Fig. \ref{fig:cube}, there is no
common point for the red, green, blue, and vertical lines. Therefore,
no stable solution in Fig. \ref{fig:cube} exists.

There also exists another way to search for stable orbits in an exact
resonance with greater resolution in the initial eccentricity. We can check
whether a stable orbit exists in the given problem directly for pairs
of $\tilde{\omega}_{\text{in}}$ and $e_{\text{in}}$ determined from
the first equation of Eqs. (\ref{system}) and depicted in
Fig. \ref{fig:zero}. The results are shown in Fig. \ref{fig:substitution}.
The red, green, and blue points correspond, respectively, to solutions of
the second, third, and fourth equation in Eqs. (\ref{system}) with substituted
$e_{\text{in}}$ obtained from the first equation of Eqs. (\ref{system}).
The collisions occur on the dashed curves. As can be seen in
Fig. \ref{fig:substitution}, the red, green, and blue lines
have no common point. Therefore, as in Fig. \ref{fig:cube},
there does not exist any stable solution for a dust particle with
$R_{\text{d}}$ $=$ 2 $\mu$m, $\varrho$ $=$ 1 g/cm$^{3}$, and
$\bar{Q}'_{\text{pr}}$ $=$ 1 captured in the exact exterior mean
motion orbital 2/1 resonance with Neptune under the action
of the PR effect, the radial solar wind, and the IGF.

We have three equations and only two unknowns. Due to the complicated
behavior of the solutions of these equations (see Fig. \ref{fig:cube} and
\ref{fig:substitution}) any dependence between the solutions is
unlikely. Therefore we can say that, most probably, no stable
solution exists for the exact resonance under the action of the PR effect,
the radial solar wind, and the IGF.

An increase of the resonant libration amplitude as shown in
Figs. \ref{fig:stabilization} and \ref{fig:oscillation}
suggests that there will be no stable solution in the sense described
in \citet{gomes95}. Such a stable solution requires a constant value
of the eccentricity (universal eccentricity) and a decrease of the resonant
libration amplitude. As can be seen in Figs. \ref{fig:stabilization}
and \ref{fig:oscillation}, the second condition is not
fulfilled for an exterior MMR with the PR effect, the radial solar
wind, and the IGF. A solution stable in the sense of \citet{gomes95}
converges to a zero amplitude of resonant libration. The zero amplitude
of resonant libration would occur also for a stable solution
in the exact resonance (see the comment after Eq. \ref{whole}).

\subsection{Non-exact resonance}
\label{sec:non-exact}

If the assumption about the exact resonance is not used, then
four unknowns remain. We use $a_{\text{in}}$, $\tilde{\omega}_{\text{in}}$,
$f_{\text{in}}$, and $f_{\text{P in}}$. Hence, from the first equation in
Eqs. (\ref{system}), we determine $e_{\text{in}}$. The non-exact
resonance will be created by a shift $\triangle$ from the exact value
of the semimajor axis $a_{\text{in}}$ $=$ $a_{\text{P}}$ $( 1 - \beta )^{1/3}$
$\left [ M_{\star} / ( M_{\star} + M_{\text{P}}) \right ]^{1/3}$
$[ p / ( p + q ) ]^{2/3}$. It is reasonable to expect that the value
of the semimajor axis for a stationary orbit in the non-exact resonance
will lie within an interval of the resonant libration of the semimajor axis.

We searched for the stationary orbits using numerous values of the shift
from the exact exterior mean motion orbital 2/1 resonance with Neptune.
If we use some value of the shift, then the remaining phase space of solutions
is three-dimensional with $\tilde{\omega}_{\text{in}}$, $f_{\text{in}}$,
and $f_{\text{P in}}$ on the axes. Sections of this phase space for nine
different longitudes of perihelion determined for $\triangle$ $=$ 0.075 AU
and $\triangle$ $=$ -0.075 AU are shown in Fig. \ref{fig:shift-p} and
Fig. \ref{fig:shift-n}, respectively. In Fig. \ref{fig:stabilization}
and Fig. \ref{fig:oscillation} it is possible to verify that these
values of the shift can lie within the interval of the resonant
libration of semimajor axis. We used a dust particle with
$R_{\text{d}}$ $=$ 2 $\mu$m, $\varrho$ $=$ 1 g/cm$^{3}$, and
$\bar{Q}'_{\text{pr}}$ $=$ 1 under the action
of the same non-gravitational effects as in
Fig. \ref{fig:cube} and Fig. \ref{fig:substitution}.
Fig. \ref{fig:shift-p} and Fig. \ref{fig:shift-n} use the same types
of lines for the same solutions of the system of equations given
by Eqs. (\ref{system}) as in Fig. \ref{fig:cube} and
Fig. \ref{fig:substitution}. For the longitudes of perihelion 0$^{\circ}$,
40$^{\circ}$, and 80$^{\circ}$, there is no solution of the fourth equation in
Eqs. (\ref{system}). Solutions of the fourth equation in Eqs. (\ref{system})
are in this case obtained close to the collisions of the planet with
the particle and such collisions do not occur for
$e_{\text{in}}$ $\apprle$ 0.2957 in Fig. \ref{fig:shift-p} and
$e_{\text{in}}$ $\apprle$ 0.2932 in Fig. \ref{fig:shift-n}.

Figures \ref{fig:shift-p} and \ref{fig:shift-n} may lead to
the idea that there is a similarity between the depicted
solutions of the equations from the system given by Eqs. (\ref{system})
for chosen values of $\triangle$ and $\tilde{\omega}_{\text{in}}$,
a similarity in the sense that the solutions of various equations
in Eqs. (\ref{system}) for chosen values of $\triangle$ and
$\tilde{\omega}_{\text{in}}$ always go side by side. If such a similarity
exists, then the solutions of the various equations depicted in
Fig. \ref{fig:shift-p} and Fig. \ref{fig:shift-n} could never have
a common point and no stationary solution could exist. This looks even
more interesting if we realize that the points with $f_{\text{P in}}$
$=$ 0$^{\circ}$ in Fig. \ref{fig:substitution} are obtained from
a figure like Fig. \ref{fig:shift-p} and Fig. \ref{fig:shift-n}
created for $\triangle$ $=$ 0 AU. In reality, the similarity is obtained
only for $\triangle$ $=$ 0 AU.

The various values of $\triangle$ in Figs. \ref{fig:shift-p} and
\ref{fig:shift-n} lead to small shifts of the depicted curves. The greatest
difference is in the plots obtained for $\tilde{\omega}_{\text{in}}$
$=$ 320$^{\circ}$. In these plots the depicted behavior of the
red curves is different for larger values of $f_{\text{in}}$.
For $\tilde{\omega}_{\text{in}}$ $=$ 320$^{\circ}$, we obtain $e_{\text{in}}$
$\approx$ 0.3388 for Fig. \ref{fig:shift-p} and $e_{\text{in}}$ $\approx$
0.3397 for Fig. \ref{fig:shift-n} from the first equation in
Eqs. (\ref{system}) and this is in both cases the value closest to the limit
value of initial eccentricity above which collisions can occur.

The shown intervals [0$^{\circ}$, 360$^{\circ}$]
for $f_{\text{in}}$ and $f_{\text{P in}}$ in Figs. \ref{fig:shift-p}
and \ref{fig:shift-n} do not allow showing the detailed behavior caused
by $\triangle$ $\neq$ 0 AU. Fig. \ref{fig:zoom} was created
in order to show a region in which these details can be seen. The region
is obtained for $\tilde{\omega}_{\text{in}}$ $=$ 200$^{\circ}$ and
various values of the shift. The same dust particle, resonance, and
non-gravitational effects are used as in Figs. \ref{fig:shift-p}
and \ref{fig:shift-n}. Therefore, the first and the last plot in
Fig. \ref{fig:zoom} correspond to the zoomed regions in the sixth plot
of Fig. \ref{fig:shift-n} and Fig. \ref{fig:shift-p}, respectively.
The region is centered on initial conditions that directly lead
to a collision for $\triangle$ $=$ 0 AU. The collision occurs
for $f_{\text{in c}}$ $\approx$ 222.91$^{\circ}$
and $f_{\text{P in c}}$ $\approx$ 62.91$^{\circ}$.
These values can be easily obtained from equations $r$ $=$ $r_{\text{P}}$,
$\cos (f_{\text{in c}} + \tilde{\omega}_{\text{in}})$ $=$
$\cos f_{\text{P in c}}$, and
$\sin (f_{\text{in c}} + \tilde{\omega}_{\text{in}})$ $=$
$\sin f_{\text{P in c}}$. For the positive values of $\triangle$
depicted in Fig. \ref{fig:zoom}, there is no collision during
the synodic period for $f_{\text{in}}$ slightly larger than
$f_{\text{in c}}$ for a given plot in Fig. \ref{fig:zoom}.
This can be easily understood since
$T_{\text{S}}$ $=$ $2 \pi q / ( n - n_{\text{P}} )$ $<$ $2 \pi / n$
for $q$ $=$ -1. For the negative values of $\triangle$, we obtain
a second collision during the synodic period for $f_{\text{in}}$
slightly smaller than $f_{\text{in c}}$.

The red, green and blue lines in Fig. \ref{fig:shift-p} and
Fig. \ref{fig:shift-n} have no common point.
Therefore no stationary solution from
the initial conditions used in Fig. \ref{fig:shift-p} and
Fig. \ref{fig:shift-n} can be obtained. It is impossible to present all
the sections of the phase space given by $a_{\text{in}}$,
$\tilde{\omega}_{\text{in}}$, $f_{\text{in}}$, and
$f_{\text{P in}}$ in a reasonable resolution in this paper.
This phase space was carefully explored for a dust particle with
$R_{\text{d}}$ $=$ 2 $\mu$m, $\varrho$ $=$ 1 g/cm$^{3}$, and
$\bar{Q}'_{\text{pr}}$ $=$ 1 captured in the exterior mean motion
orbital 2/1 resonance with Neptune under the action of the PR effect,
the radial solar wind, and the IGF and no stationary orbit was found.

\section{Conclusions}
\label{sec:conclusions}

From a near canonical form of the equations of motion,
we derived averaged resonant equations for a circumstellar
dust particle captured in an MMR with a planet in a circular orbit under
the action of given non-gravitational effects in general form. The averaged
resonant equations were also obtained/confirmed using Lagrange's
planetary equations. Non-gravitational effects for which the secular
variations of the orbit depend on the spatial orientation of the orbit
have been considered. This enabled us to investigate the motion of a dust
particle in an MMR under the action of the PR effect, the radial stellar wind,
and the interstellar wind. The analytically and numerically calculated
averaged time derivatives of the non-canonical resonant variables
are in excellent agreement.

From the numerical solutions of the equation of motion,
the evolutions in the $kh$ plane can be obtained. Using
the evolutions in the $kh$ plane, the types of orbits can be easily
determined. For an exterior mean motion orbital 2/1 resonance, all types
of orbits correspond to libration centers of the conservative problem.

From the averaged resonant equations, one can obtain
a system of equations characterizing possible
stationary solutions in a given problem. Using
that system we analytically showed all stationary solutions
must correspond to orbits which are stationary in
the interplanetary space after averaging over the synodic
period. We found that the stationary obits in the exact resonance
repeat their evolution periodically after the synodic period.
In such a configuration, the orbit is stable.

The obtained system can be numerically solved in
order to find stationary solutions. An exact
resonance yields a phase space of solutions which has two
dimensions. Since the number of remaining equations is three and
some dependence between the solutions is unlikely, most probably no
stationary solutions in the exact resonance exist.
For a non-exact resonance, the phase space of solutions has four
dimensions. Solutions of all three remaining equations in
this phase space are very similar except for the behavior close to
collisions. This similarity does not enable obtaining a stationary
solution. For the used test particle in the non-exact exterior 2/1
resonance under the action of the PR effect, the radial solar wind
and the interstellar wind, no stationary solution is located
in the four-dimensional phase space of solutions.

\begin{acknowledgements}
I would like to thank the referees of this paper for their useful
comments and suggestions.
\end{acknowledgements}

\end{document}